# Impact of Semiconductor Band Tails and Band Filling on Photovoltaic Efficiency Limits

*Joeson Wong*[ǂ], *Stefan T. Omelchenko*[ǂ], *and Harry A. Atwater*[*]

Department of Applied Physics and Materials Science, California Institute of Technology,

Pasadena, CA-91125, USA

ǂ These authors contributed equally

*Harry A Atwater (haa@caltech.edu)

## MAIN TEXT

Since the seminal work of Shockley and Queisser, assessing the detailed balance between absorbed and emitted radiative fluxes from a photovoltaic absorber has been the standard method for evaluating solar cell efficiency limits[1–3]. The principle of detailed balance is one dictated by reciprocity and steady state, so that photons can be absorbed and emitted with equal probability. This basic principle has also been extended to evaluate the effects of multiple junctions[4,5], hot carriers[6,7], nanostructured geometries[8,9], multiexciton generation[10,11], sub-unity radiative efficiency[12] and many other solar cell configurations and nonidealities to estimate limiting efficiencies via modifications to the detailed balance model.

Another important modification to the Shockley-Queisser model is to examine the assumption of an abrupt, step-like onset of the densities of electronic states and absorption coefficient. Specifically, it has long been recognized from spectroscopic measurements of semiconductors that band edges are often not abrupt and that the density of states and absorption functions can be characterized by a band tail. This was first recognized by Urbach[13], who found the absorption coefficient for a variety of materials below their bandgaps to be characterized by an exponential tail:

$$\alpha\left(E < E_g\right) = \alpha_0 \exp\left(\frac{E - E_g}{\gamma}\right) \tag{1}$$

where $\alpha_0$ is the absorption coefficient value at the energy of the bandgap, $E_g$ is the bandgap of the material, and $\gamma$ is referred to as the Urbach parameter, which describes the rate at which the absorption coefficient goes to zero. The magnitude of the Urbach parameter can be influenced by impurities and disorder and is typically attributed to fluctuations in the electrostatic potential within a semiconductor. Urbach tails have been observed in a wide range of absorber materials including amorphous, organic, perovskite, and II-VI, III-V and group IV semiconductors[14–20].

While the Urbach exponential tail is the most prominent functional form observed for band tail states, other forms such as Gaussian band tails have been reported, and different functional forms have been attributed to the underlying physics of those systems[21]. In most cases, the band tail can be characterized by an exponential with an argument raised to some power.

Recent detailed balance analyses have also suggested how this important effect, *i.e.* a departure from a step-like absorbance spectra can also degrade the limiting efficiency of solar cells[2,20,22–25]. However, a key element missing from previous analyses of photovoltaic efficiency is the effect of band filling for semiconductors with nonabrupt band edges, wherein the electron-hole quasi-Fermi level splitting can thereby modify the absorption spectrum, and therefore the radiative emission spectrum as well. This voltage-dependent absorption effect was first recognized by Parrott[26] as being necessary to make the detailed balance formulation self-consistent. Perhaps the most intuitive description of why this is necessary is found by examining the generalized Planck's law[27]:

$$S_{PL}(E) = a(E)S_{bb}(E,\Delta\mu) \tag{2}$$

where

$$S_{bb}(E,\Delta\mu) = \frac{2\pi}{h^3c^2}\frac{E^2}{\exp\left(\frac{E-\Delta\mu}{kT}\right)-1} \tag{3}$$

and $S_{PL}(E)$ describes the luminescence flux, $a(E)$ is the absorbance, $E$ is the photon energy, $\Delta\mu$ is the quasi-Fermi level splitting, $h$ is Planck's constant, and $c$ is the speed of light. A clear singularity occurs at $E = \Delta\mu$, which is typically ignored in detailed balance calculations because for a step-like absorbance function, we have $E \geq E_g > \Delta\mu$. As a result, the -1 in the denominator is neglected and Boltzmann statistics are assumed. Clearly, the situation must change if we consider energies $E < E_g$, as is the case when band tails affect the luminescence spectrum. In this case, the absorptivity must be modified such that $a(E = \Delta\mu) = 0$, and in general the absorption coefficient is occupation dependent:

$$\alpha(E, \Delta\mu) = \alpha_{0K}(E)(f_v - f_c) \tag{4}$$

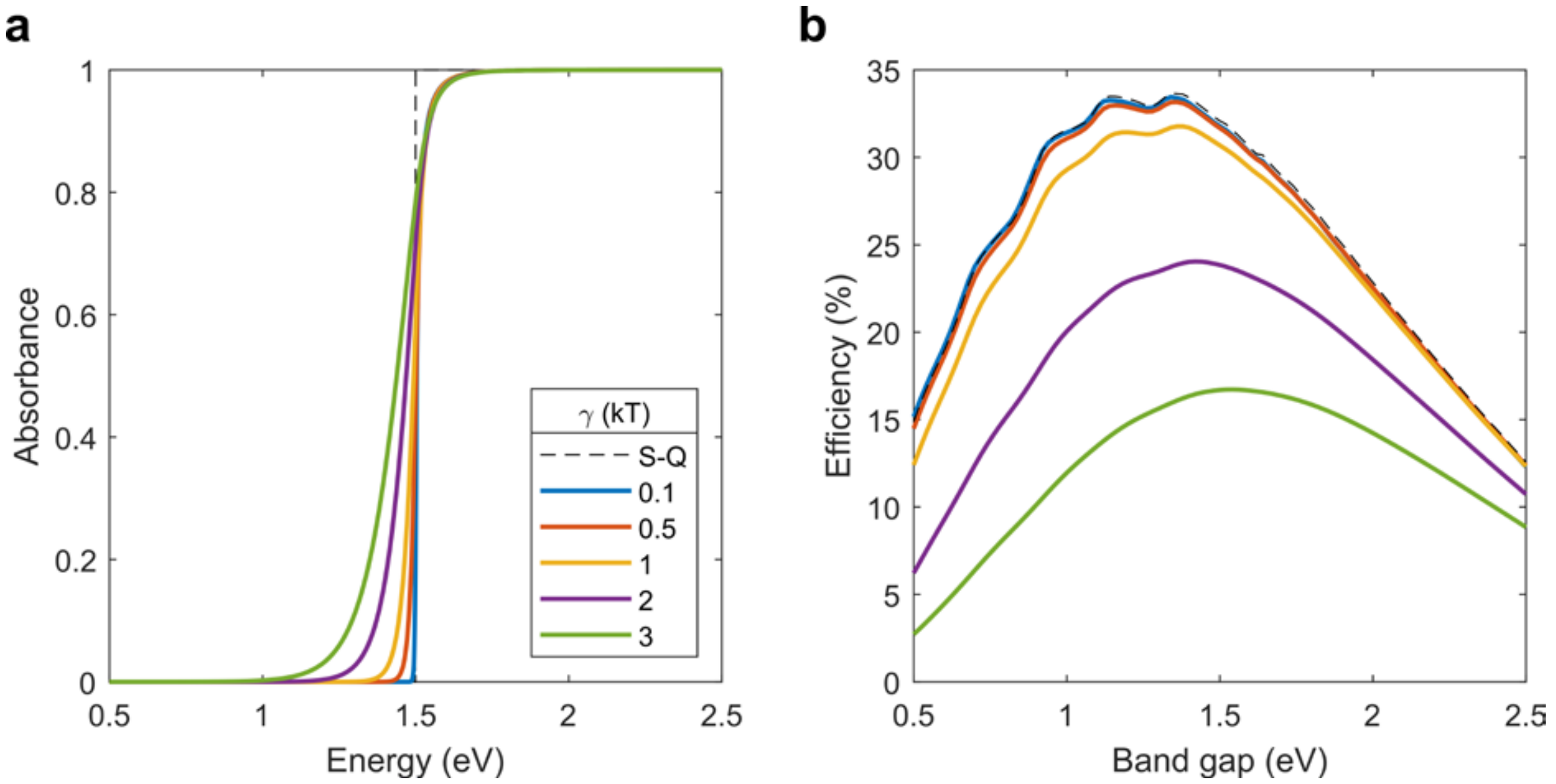

**Figure 1**. **Effects of band tailing on photovoltaic limiting efficiencies**: **(a)** The spectral absorbance of a photovoltaic cell with a bandgap of $E_g = 1.5$ eV and a thickness $\alpha_0 L = 1$ plotted for various Urbach parameters ($\gamma$) in units of $kT$. The dashed line represents the step function absorbance typically used in the Shockley-Queisser (S-Q) limit. **(b)** The detailed balance efficiencies as a function of the bandgap energy. Different colored lines correspond to different Urbach parameters, with the coloring scheme equal to the legend shown in (a).

Where $a_{0K}(E)$ is the absorption coefficient without band-filling and $(f_v - f_c)$ is the band-filling factor[21,27,28]. This contribution of band-filling has also been recognized in experiments as being necessary to accurately fit photoluminescence spectra under high level injection[29,30]. We suggest that this contribution is also important for systems with large band tails, and as an example, we have used this modified reciprocity relation to fit the electroluminescence spectrum of a-Si:H which the Rau reciprocity relation[3,31] was previously unable to fit completely (see Figure S1).

**Photovoltaic Efficiency Limit for Semiconductors with Band Tails**

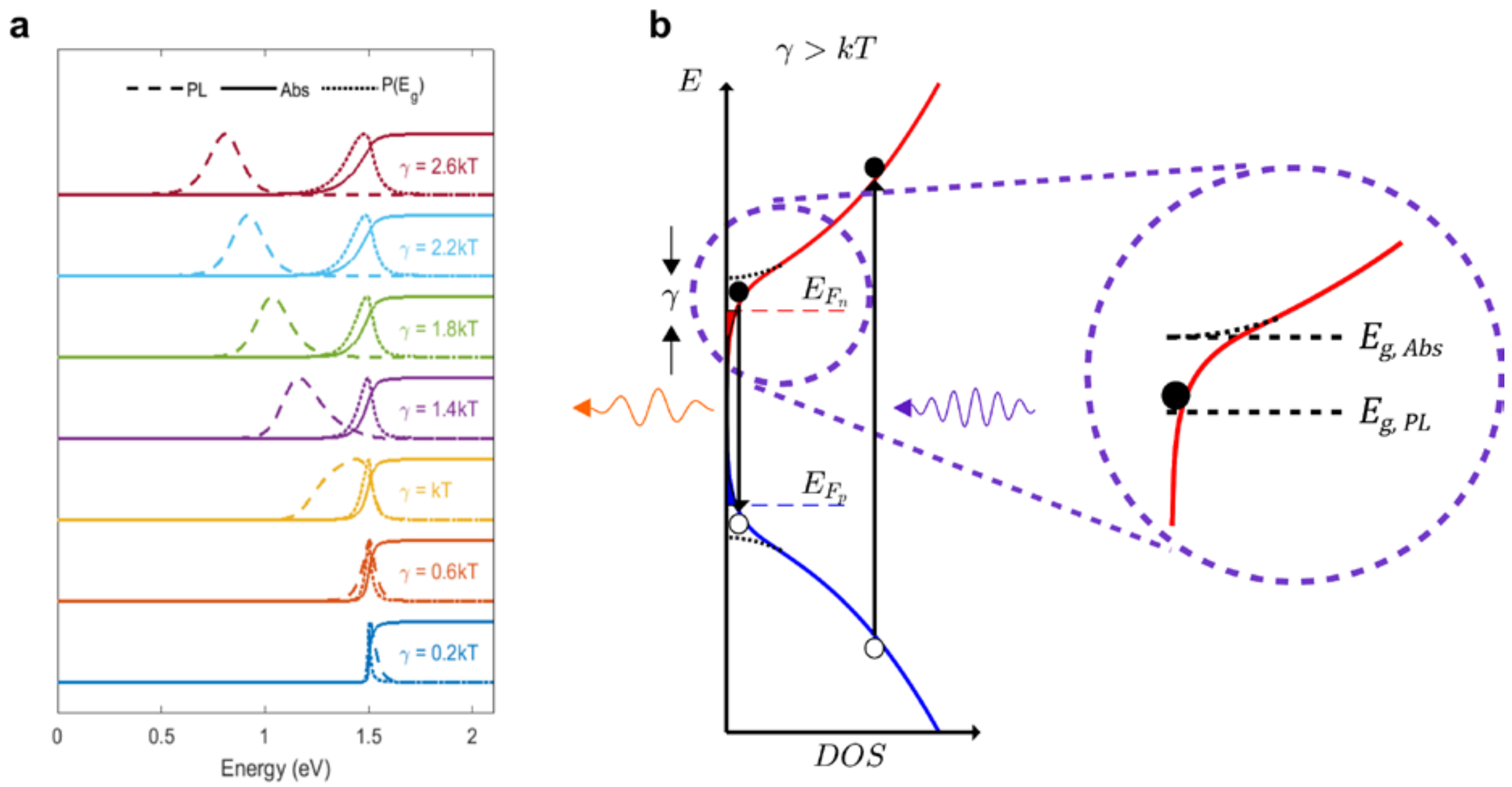

**Figure 2. Effects of band tail states on photoluminescence**: **(a)** The normalized spectral photoluminescence (dashed line) of a photovoltaic cell operating at the radiative limit under 1 sun AM 1.5G illumination for increasing Urbach parameter ($\gamma$) with an offset included for clarity. The corresponding absorbance (solid line) and effective distribution of bandgaps (dotted line) is also plotted, where they are normalized to their peak value. **(b)** Schematic depiction of the density of states profile along with carrier excitation and recombination; $\gamma$ describes the effective width of the band tail. For small band tailing ($\gamma < kT$), the effect of band tailing is to simply broaden the luminescence peak. For systems with large band tailing ($\gamma > kT$), the luminescence shifts to energies below the nominal absorption band edge.

By using the generalized Planck's law (Eqn. 2) and accounting for band filling (Eqn. 4), we can calculate the detailed balance limit for photovoltaic efficiency with band tails in the radiative limit (See Section S1 and S2 in the Supplementary Information). In Figure 1, we consider the case of a band tail parameterized as an exponential Urbach tail and analyze the effects of varying the Urbach parameter. While the spectral response of this modified absorbance appears to be similar to the step function response originally used by Shockley and Queisser (black dashed line), the maximum achievable efficiency drops rapidly from the Shockley-Queisser limit for Urbach parameters larger than the thermal energy, $kT$. These effects are relatively insensitive to the choice of bandgap and thickness (see Figure S2 and S3), and Figure S1 suggests the efficiency drop is primarily due to a voltage loss mechanism.

To analyze the cause of the voltage loss, we examine the luminescence spectrum by using Eqn. 2 and plot these spectra for various Urbach parameters (Figure 2a). In addition, we plot the distribution of bandgaps $P(E_g) = \partial_E Abs|_{E=E_g}$ proposed by Rau et al.[3] recently, which generalizes the definition of the photovoltaic bandgap for arbitrary absorbance spectra. While the luminescence spectrum is narrow and overlaps significantly with the absorption edge for $\gamma < kT$, this is not the case for $\gamma > kT$. In this limit, the luminescence spectrum is significantly broadened and shifts away from the absorption band-edge and suggests the definition of a second bandgap, defined by the luminescence spectra. This idea is schematically depicted in Figure 2b, where the absorption bandgap is defined as before, i.e. $P(E_{g,Abs}) = \max(P_{E_g})$, while the second bandgap $E_{g,PL}$, is defined by the luminescence spectra $S_{PL}$. We note that this analysis is modified significantly with the inclusion of band-filling effects, which we describe in Section S3 and S4 and Figures S4 and S5 of the Supplementary Information. We also observe the effects of broadening followed by luminescence spectral shifts in band tails parameterized by a Gaussian band tail (Figure S6), where the onset of efficiency loss occurs at approximately $\gamma = 2kT$ instead.

**Generalized Voltage Loss for Semiconductors with Nonabrupt Band Edges**

The similarity between the effects of broadening followed by luminescence shifting for increasing band tail energies suggests a general picture for the voltage loss mechanism, for any band tail functional form. A general trend is the observation of a Stokes shift, i.e. the shift between the absorbance and luminescence spectra, that occurs precisely at the onset of efficiency loss.

However, it is unclear whether the voltage loss is just directly proportional to the observed Stokes shift $\Delta E_g$.

To develop an understanding of this loss mechanism, we consider a simpler absorbance spectrum as a two bandgap model, represented by the sum of two step function absorbances:

$$a(E) = a_1\theta(E - E_{g,1})\theta(E_{g,2} - E) + a_2\theta(E - E_{g,2}) \tag{5}$$

Here, $a_{1,2}$ is the sub-gap and above-gap absorbances respectively, while $\theta(E - E_g)$ is the Heaviside step function, typically considered in the SQ analysis. The photovoltaic bandgap, i.e. that defined by absorption, is given by $E_{g,2}$, while $E_{g,1}$ defines the luminescence bandgap. The SQ limit is recovered in the limit that $E_{g,1} \to E_{g,2}$ or $a_1 \to 0$. By varying $a_1$ and $E_{g,1}$ and fixing $E_{g,2}$ to 1.34 eV and $a_2 = 1$, we can analyze the effects of this simple model as we deviate from the SQ limit (see Section S5 and Figure S7 for more details). Interestingly, we find qualitatively similar effects of voltage and efficiency loss in this absorbance model compared to the full effects of the Urbach band tail, albeit parametrized by $a_1$ and $E_{g,1}$ instead of the Urbach parameter $\gamma$. However, we also find that the quantitative bandgap-voltage relation can be significantly affected by the actual functional form used to more accurately model the band tail state distribution, as illustrated in Fig. 3.

Non-abrupt band-edge absorbances can be mapped onto the two bandgap model and therefore there is a general relation that explains the voltage loss mechanism for any absorbance spectrum given by

$$\Delta V_{oc,rad} = \frac{kT}{q}\log\left(\frac{\bar{a}_{SG}}{\bar{a}_{AG}}\exp\left(\frac{\Delta E_g}{kT}\right) + 1 - \frac{\bar{a}_{SG}}{\bar{a}_{AG}}\right) \tag{6}$$

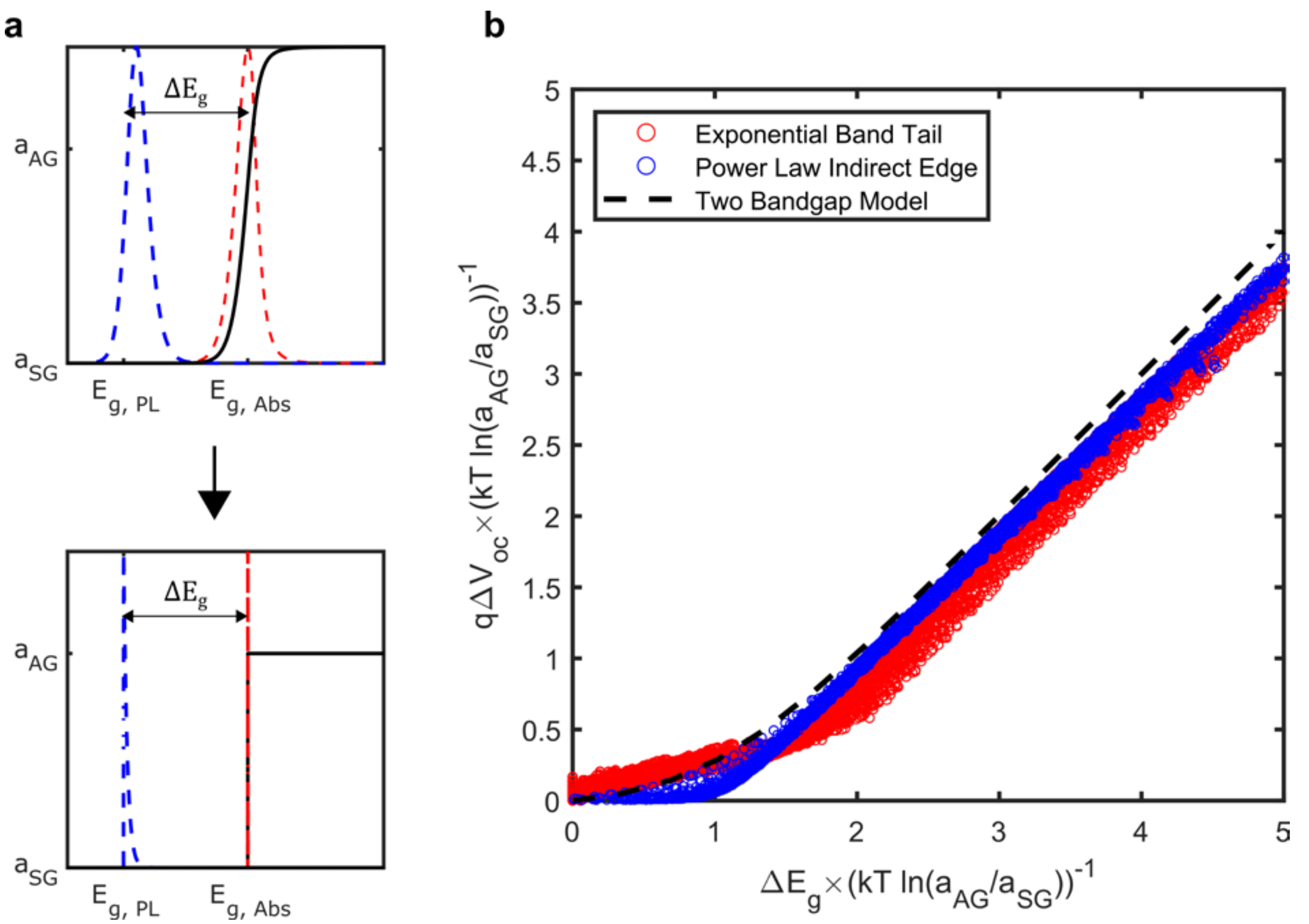

**Figure 3. Generalized voltage losses parametrized as a two-bandgap absorber: (a)** Schematic depiction representing a general absorbance and luminescence spectrum as a simpler two-bandgap step function absorbance. Black solid lines are the absorption spectrum, whereas the red dashed line corresponds to the bandgap distribution $P(E_g)$. Blue dashed lines correspond to the luminescence spectrum $S_{PL}$. Typically, $a_{SG} \ll 1$, which is not visible on a linear scale but still contributes to the luminescence spectra due to carrier thermalization. **(b)** Calculated voltage loss $\Delta V_{oc} = V_{oc,SQ}(E_{g,Abs}) - V_{oc,rad}$ versus observed bandgap shift $\Delta E_g = E_{g,Abs} - E_{g,PL}$, normalized to the energy scale $kT\ln(a_{AG}/a_{SG})$. Every plotted point corresponds to a different absorption spectrum, with $V_{oc,rad}$ calculated using the complete absorption spectra and the full reciprocity relations. The dashed line represents the two bandgap model, i.e. Eqn. 6, where we have chosen $\bar{a}_{SG}/\bar{a}_{AG} = 0.1$.

Where $\bar{a}_{SG}$ is the weighted sub-gap absorbance, $\bar{a}_{AG}$ is the weighted above-gap absorbance, and $\Delta E_g = E_{g,Abs} - E_{g,PL}$ describes the observed Stokes shift between the absorption and luminescence (see definitions in section S6 of the SI). Here, $\Delta V_{oc,rad}$ is a voltage loss due *purely* to the non-abruptness of the absorption spectrum, for a semiconductor with assumed unity radiative efficiency. More generally, although non-radiative losses parametrized by a non-unity external radiative efficiency have not been accounted for (see some discussion of radiative efficiency effects in Section S7 of the SI), band edge non-abruptness by itself can contribute significantly to voltage loss. Indeed, Eqn. 6 results in no net voltage loss as $\Delta E_g \rightarrow 0$, and suggests that a finite Stokes shift should be directly correlated to a voltage loss. The magnitude of the voltage loss is scaled by the ratio $\bar{a}_{SG}/\bar{a}_{AG}$, and clearly as the ratio approaches 0 or 1, Eqn. 6 recovers the appropriate losses of 0 and $\Delta E_g/q$, respectively.

To observe whether this two bandgap model can quantitatively describe the more complex band edge functional forms seen in experiments, we choose appropriate definitions for $\bar{a}_{SG}, \bar{a}_{AG}$, and $E_{g,PL}$ and use Rau's definition for $E_{g,Abs}$ (see more details in Section S6, Section S8, and Figure S9). We consider both power law band-edges, as a parametrization of indirect band-edges, as well as exponential band tails. We find reasonable qualitative agreement but quantitative disagreement between the calculations utilizing the full absorbance spectra and that given by Eqn. 6 (Figure 3b), suggesting that the two-bandgap model is a reasonable first-order representation of the voltage loss mechanism, but importantly, consideration of the actual band tail functional form yields more accurate results. Furthermore, we find that the dimensionless parameter $\xi = \Delta E_g/(kT \log(\bar{a}_{AG}/\bar{a}_{SG}))$ describes the physical regime of voltage loss. Generally, for $\xi < 1$, voltage loss is minimal since the emission spectrum can be considered as simply a broadening of a single photovoltaic bandgap. In this regime, the efficiency penalty is negligible and generally the detailed balance efficiency limit for $E_{g,Abs}$ can be achieved given sufficient absorption above the photovoltaic bandgap. However, for $\xi > 1$, it is appropriate to define a second bandgap given by the emission spectrum (Figure 3a), resulting in a substantial voltage and efficiency penalty due to additional thermalization losses. Thus, the tuning of the band tail parameter $\gamma$ merely sweeps through different values of $\xi$, and we find that $\xi > 1$ is equivalent to $\gamma > kT$ in the case of an Urbach tail (see Section S5 in SI). The discrepancy in Eqn. 6 for large $\xi$ can be attributed to neglecting higher order terms (see Section S6 in SI).

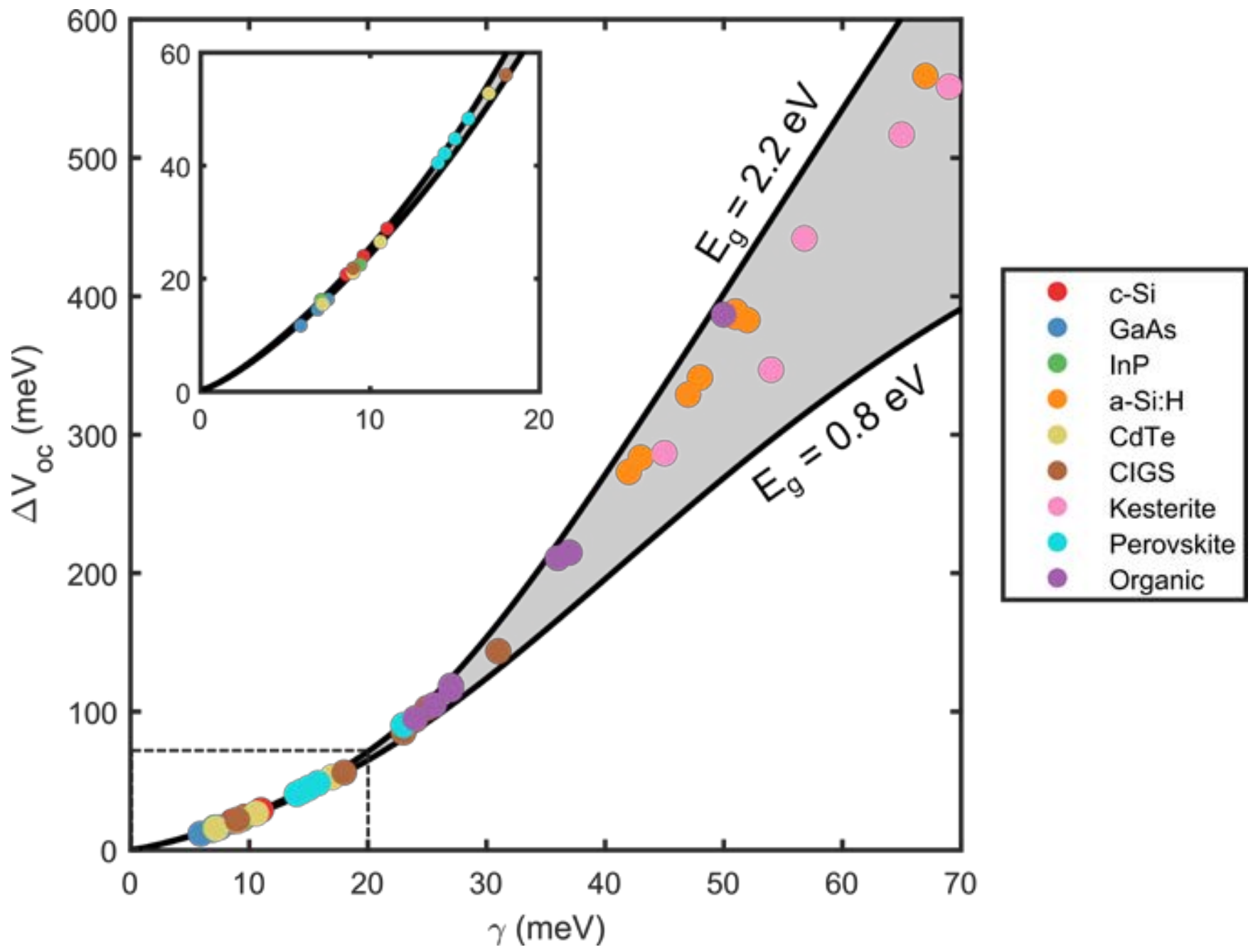

**Figure 4. Voltage loss due to a nonabrupt band-edge**: Expected open circuit voltage loss as a function of the observed Urbach parameter $\gamma$, plotted for different materials. The top and bottom black solid lines represent the calculated voltage loss assuming a bandgap of 2.2 and 0.8 eV, respectively. The gray area in between represents the voltage loss expected for bandgap values in between 0.8 and 2.2 eV, which correspond to most of the materials considered for photovoltaics. The colored data points indicate the expected voltage loss for an experimentally-measured Urbach parameter, The dashed line corresponds to the region of the inset, where the voltage loss is minimal and approximately the same irrespective of bandgap.

The correlation between the magnitude of the bandgap shift (i.e. $\Delta E_g$) and open circuit voltage has already been recognized in the organic photovoltaics literature, where the presence of low energy charge transfer states generally results in cells with a lower voltage and efficiency[20,32–35].

Here, we have developed a unified picture with an arbitrarily-shaped band tail and by explicitly including band-filling effects, for both large and small band tails, the voltage loss mechanism can be qualitatively captured with a simple two bandgap model. In addition, by extracting the weighted absorbance ratios, $\bar{a}_{AG}/\bar{a}_{SG}$, we can estimate the voltage losses in the radiative limit using Eqn. 6. We therefore suggest that any radiative transition below the photovoltaic absorption edge $E_{g,Abs}$, measurable in luminescence measurements, should result in an efficiency penalty. This efficiency penalty can be viewed as either stemming from a voltage penalty, due to carrier thermalization within the band tails, or equivalently interpreted as being due to incomplete absorption at the lower energy transition.

To emphasize the implications of these results for various photovoltaic technologies, we have calculated the predicted voltage losses due to a nonabrupt band-edge for several different material systems and plotted experimentally-measured values for these in Figure 4 (See Table S1 for references and individual datapoints)[14,20,22]. Since Urbach parameters are much more commonly reported than both high sensitivity EQE and EL spectra, we have used the observed Urbach parameters to calculate the voltage loss directly rather than through Eqn. 6. As expected, we find that semiconductors with large band tails ($\gamma > kT$) or equivalently, large Stokes shifts ($\Delta E_g \gg kT$) have a substantially modified maximum achievable $V_{oc}$, which should be assessed when examining their efficiency potential (e.g., CIGS, a-Si, kesterites, and OPVs). It should be noted that a more accurate calculation can be made by using the directly measured EQE and EL spectra for a given device.

The analysis presented here should be applicable to any system with nonabrupt band-edges that obeys the optoelectronic reciprocity relations and should be employed to evaluate the radiative limits on the open circuit voltage. We demonstrated here that the voltage dependence of the absorbance or EQE, specifically via band filling, must be included to self-consistently apply the generalized Planck's law for semiconductors. We also suggest that in order to accurately estimate efficiency limits, the abruptness of the band-edge should be experimentally characterized by measuring both the absorption and luminescence spectra of photovoltaic materials, and in a completed photovoltaic device, photocurrent and electroluminescence spectra should be used to assess the effects of transport on the reciprocity relations. The magnitude of the voltage loss can then be estimated directly from the spectroscopic measurements by applying reciprocity relations. Additional experimental details and nonidealities for a given photovoltaic material or device may modify the maximum efficiency potential even further, such as reduction in the external radiative efficiency or the finite mobility of charge carriers. However, our analysis suggests the important role that band edge abruptness and band filling can play in defining the limit on open circuit voltage and efficiency potential of emerging and established photovoltaic materials.

ASSOCIATED CONTENT

**Supporting Information**

Reciprocity relations background; calculation details; discussion of band-filling contribution and modified *J-V* characteristics; discussion of two bandgap model; derivation of generalized voltage loss mechanism and different band-edge parametrizations; effects of different bandgaps, Urbach parameters, and thickness on photovoltaic figures of merit; effects of sub-unity radiative and quantum efficiencies; Analysis of Gaussian tail distributions.

**Author Contributions**

ǂThese authors contributed equally. J.W. and S.T.O. developed the ideas, formulism, and performed the calculations for all the results shown in this work. H.A.A supervised over all the calculations and analysis. J.W. wrote the manuscript, with input from S.T.O and H.A.A. All authors contributed to the discussion and interpretation of results, as well as the presentation and preparation of the manuscript.

**ORCID**

Joeson Wong: 0000-0002-6304-7602

Stefan T. Omelchenko: 0000-0003-1104-9291

Harry A. Atwater: 0000-0001-9435-0201

**Notes**

Views expressed in this Viewpoint are those of the authors and not necessarily the views of the ACS. The authors declare no competing financial interest.

ACKNOWLEDGMENT

This work is part of the 'Photonics at Thermodynamic Limits' Energy Frontier Research Center funded by the U.S. Department of Energy, Office of Science, Office of Basic Energy Sciences under Award Number DE-SC0019140. J.W. acknowledges additional support from the National Science Foundation Graduate Research Fellowship under Grant No. 1144469.

Supporting Information for

# Impact of Semiconductor Band Tails and Band Filling on Photovoltaic Efficiency Limits

*Joeson Wong*ǂ, *Stefan T. Omelchenko*ǂ, *and Harry A. Atwater**

Department of Applied Physics and Materials Science, California Institute of Technology, Pasadena, CA-91125, USA

* Corresponding author: Harry A Atwater (haa@caltech.edu)

ǂ These authors contributed equally

**Table of Contents:**

## Section S1. Optoelectronic Reciprocity Relations

The connection between absorption and emission has been known for quite some time. Kirchhoff in 1860[1] is often cited as being the first to recognize the relation between the two processes, noting that the absorption and emission probability of a photon must be equal, i.e. $a(E) = e(E)$, through arguments of thermal equilibrium. A surface with $e(E) = 1$ for all energies is known as a perfect black body. However, the precise spectral dependence of a perfect black body emitter was not derived until Planck did so in 1906[2]. He theorized a cavity with perfectly absorbing walls filled with a gas of photons with a small hole that would leak out a spectral flux characteristic of a black body:

$$I_{bb}(E) = E \times S_{bb}(E) = \frac{2\pi}{h^3c^2}\frac{E^3}{\exp\left(\frac{E}{kT}\right) - 1} \tag{1}$$

which relates the temperature of a black body to its spectral characteristics, often referred to as *thermal* radiation. Here, $S_{bb}(E)$ is the energy-resolved photon flux per unit area per unit time of black-body radiation and $I_b$ refers to the spectrally resolved intensity of the radiation. The above expression can be generalized to non-black bodies by combining it with Kirchhoff's law:

$$I(E) = a(E)I_{bb}(E) \tag{2}$$

To form a general law for thermal radiation with *surfaces* characterized by an absorptivity. In an analogous manner, van Roosbroeck and Shockley[3] generalized Planck's law to semiconductors and related the absorption coefficient ($\alpha$) to the *internal* photon emission rate per unit *volume*:

$$R(E) = 4n_r^2(E)\alpha(E)S_{bb}(E) = \frac{8\pi n_r^2}{h^3c^2}\frac{E^2\alpha(E)}{\exp\left(\frac{E}{kT}\right) - 1} \tag{3}$$

which holds for systems *at thermal equilibrium*. It was not until Lasher and Stern[4] considered the situations of *spontaneous emission* were the above expressions further generalized to include *non-equilibrium, steady-state* conditions in terms of the quasi-Fermi level splitting $\Delta\mu$, which is exactly equal to the chemical potential of the photon in a spontaneous emission process:

$$R(E) = \frac{8\pi n_r^2}{h^3c^2}\frac{E^2\alpha(E)}{\exp\left(\frac{E-\Delta\mu}{kT}\right) - 1} \tag{4}$$

Here, we note that that this expression is valid only when quasi-thermal equilibrium holds, where exactly two different quasi-Fermi levels accurately describe the energy dependence of the two separate populations of electrons and holes (e.g. after the electron-electron interactions subsequent to the excitation of carriers, the carriers will be distributed according to the Fermi-Dirac distribution), resulting in a single quasi-Fermi level splitting $\Delta\mu$. We assume this to be true in the case of the carriers in the band tails described here with the carriers above the respective band edges. For example, in the case of band tails caused by some ensemble of defects, an impurity band may be formed. If this impurity band is several $kT$ away from the band edges, the electrons in this band would likely thermalize amongst themselves, forming a separate quasi-Fermi level. Therefore, these relations would need to be modified to include this effect.

Wurfel[5] then generalized the Lasher-Stern relation to an *external* flux of radiative emission from a semiconductor surface:

$$S_{PL}(E) = a(E)S_{bb}(E, \Delta\mu) \tag{5}$$

where

$$S_{bb}(E, \Delta\mu) = \frac{2\pi}{h^3c^2}\frac{E^2}{\exp\left(\frac{E-\Delta\mu}{kT}\right)-1} \tag{6}$$

is the spectral flux of a photon gas with chemical potential $\Delta\mu$ and temperature $T$. Here, $E$ is the energy of the emitted photon, $k$ is the Boltzmann constant, $h$ is Planck's constant, and $c$ is the speed of light. Wurfel's expression, with a relation that connects the absorbance to the absorption coefficient (e.g. through the Beer-Lambert law of $a(E) = 1 - \exp(-\alpha L)$ or more complex light-trapping geometries) suggests a complete set of self-consistent expressions that connect *external* properties (e.g absorbance, external luminescence) of the semiconductor to its *internal* properties (e.g. bandgap, absorption coefficient, quasi-fermi level splitting, internal luminescence). External properties are therefore geometry dependent and can be carefully engineered from the internal properties using photonic design. Moreover, external properties are typically the only properties that are experimentally accessible.

We note that the above expression has an apparent divergence at $E = \Delta\mu$. The resolution requires including an occupation factor in the absorption coefficient:

$$\alpha(E) = \alpha_{0K}(E)(f_v - f_c) \tag{7}$$

Where $f_v$ and $f_c$ are the occupation for the holes and electrons, respectively. In the case of a semiconductor with equal effective mass for the holes and electrons and described by a parabolic dispersion, the occupation factor has a simple form:

$$f_v - f_c = \tanh\left(\frac{E-\Delta\mu}{4kT}\right) \tag{8}$$

While real systems may have more complex occupation factors (typically not representable analytically due to a fairly complex band structure), we note that $f_v - f_c$ is generally a function with limiting values from -1 to 1 with a value of zero at $E = \Delta\mu$, which is captured by the simple expression above. For simplicity and to capture the physics of the band filling irrespective of other materials properties, we use the simple expression above when calculating band filling effects.

It was suggested more recently by Rau[6] that the principle of optical reciprocity can be further generalized to an *optoelectronic* reciprocity by including the serial collection/injection with Donolato's theorem[7] to describe photovoltaic cells and LEDs:

$$S_{EL}(E) = EQE(E)S_{bb}(E, \Delta\mu) \tag{9}$$

Where $EQE(E) = a(E) \times IQE(E)$ and describes the process of absorbing a photon with probability $a(E)$ with a subsequent collection probability of $IQE(E)$. Thus, the LED quantum efficiency $Q_{LED}(E) = \eta_{inj}(E) \times e(E)$ is a detailed balance pair with the photovoltaic quantum efficiency, taking the injection and collection efficiencies to be detailed balance pairs. We note that while the above generalized Planck's law (Wurfel's expression) holds quite generally by any system that can be characterized by two distinct quasi-Fermi levels and a thermodynamic temperature, Rau's reciprocity relation strictly holds only in systems where carrier transport under illumination is well modelled as a linear perturbation of thermal equilibrium (qualitatively, the law of superposition in the current-voltage curves needs to hold). We also note that previous examples of using optoelectronic reciprocity for photovoltaic analysis (e.g. modified detailed balance models) has often approximated the black-body flux as:

$$S_{bb}(E, \Delta\mu) \approx S_{bb}(E, 0) \exp\left(\frac{\mu}{kT}\right) = \left(\frac{2\pi}{h^3 c^2} \frac{E^2}{\exp\left(\frac{E}{kT}\right) - 1}\right) \exp\left(\frac{\mu}{kT}\right) \tag{10}$$

While the above expression has no singularities and generally results in numerically accurate results for most systems of interest (e.g. idealistic systems with $a\left(E \leq E_g\right) = 0$ will generally have $\frac{E-\mu}{kT} \gg 1$), De Vos and Pauwels[8] noted the subtle differences this approximation has in analyzing entropy generation in the detailed balance limit. We show in this paper that accounting for band filling effects has qualitative and quantitative differences on the luminescence spectra of semiconductors with significant band tailing, which we emphasize in Figure S1 with a-Si:H as an example. Therefore, we use the full expression above without any approximations.

**Section S2. Modified Detailed Balance Limit Calculations**

With the above expressions of optoelectronic reciprocity in hand, we can assemble a modified detailed balance model for solar cells that account for carrier generation, extraction, and recombination:

$$\int EQE(E, \Delta\mu) S(E) dE = \frac{\int EQE(E, \Delta\mu) S_{bb}(E, \Delta\mu) dE}{\eta_{ext}(\Delta\mu)} + \frac{J(\Delta\mu)}{q} \tag{11}$$

Where the left-hand side describes carrier injection (e.g. from sunlight or other light source) and the right-hand side describes carrier extraction, either through radiative recombination, non-radiative recombination (parametrized by $\eta_{ext}(\Delta\mu)$), or usefully as carrier collection ($J(\Delta\mu)/q$). In steady state, these populations must be balanced. In our analysis in the main text, we consider the modified detailed balance expression in the radiative limit i.e. $\eta_{ext} = 1$, $\Delta\mu = qV$, $EQE(E, \Delta\mu) = a(E, \Delta\mu)$ (see Section S8 for a short analysis on non-unity radiative or collection efficiency), with absorptivity described by a Beer-Lambert expression

$$a(E) = 1 - \exp(-2\alpha L) \tag{12}$$

With a perfect back reflector and perfect anti-reflection coating to describe the optical configuration. To parametrize the band edge density of states, we take inspiration from Katahara and Hillhouse[9] and convolve a sub-gap exponential density of states with a parabolic density of states above the bandgap, giving:

$$\alpha_{0K}(E) = \alpha_0 \sqrt{\frac{\gamma}{kT}} G\left(\frac{E - E_g}{\gamma}\right) \tag{13}$$

With

$$G(x) = real\left(\frac{1}{2\Gamma\left(1 + \frac{1}{\theta}\right)} \int_{-\infty}^{\infty} \exp(-|x'|^{\theta}) \sqrt{x - x'} dx'\right) \tag{14}$$

And the simplified expression above (Eqn. 8) to account for band filling. Here, $\gamma$ is the energy width parameter (i.e. the Urbach parameter, for $\theta = 1$). $E_g$ is the bandgap, $\Gamma$ is the Gamma function, $\alpha_0$ scales the absorption coefficient (i.e. $\alpha(E = E_g) = \alpha_0\sqrt{\pi\gamma/16kT}$), and $\theta$ describes the power of the sub-gap exponential distribution. Our expression has an extra factor of $\sqrt{kT}$ compared to the Katahara model, where $kT$ is the thermal energy, so that $\alpha_0$ has the usual units of absorption coefficient. Using a simple piecewise continuous function for the absorption coefficient above and below the gap yields similar results, as long as the absorption coefficient below the gap is still modeled as an Urbach tail. Thus, for a given set of materials parameters (e.g. $\alpha_0 L$, $\gamma$, $E_g$) and a specific voltage $V = \Delta\mu$, we can calculate the appropriate absorption coefficient and consequently the absorption and luminescence characteristics. The current-voltage curve of the photovoltaic cell in the detailed balance limit is then calculated using Equation 11. Specific figures of merit can then be extracted from the current-voltage curves.

### Section S3. Band filling Contribution to Photoluminescence

In general, we are interested in the contribution of including the band filling on the luminescence spectrum of a semiconductor with significant band tails. Let us examine the case where we are weakly absorbing, which is generally true in the spectral region of a band tail. In this limit, we can take $a \approx 2\alpha L$, where we assume a planar system with a perfect mirror and a perfect antireflection coating as above. In this case, the external luminescence flux by reciprocity becomes

$$S_{PL}(E, \Delta\mu) = a(E, \Delta\mu) S_{bb}(E, \Delta\mu) \approx 2\alpha(E, \Delta\mu) L S_{bb}(E, \Delta\mu) \tag{15}$$

For systems with intrinsic doping and equal effective masses, we have $\alpha(E, \Delta\mu) = \alpha(E, 0)(f_v - f_c) = \alpha(E, 0)\tanh\left(\frac{E-\Delta\mu}{4kT}\right)$. To see this, note that generally speaking,

$$f_v - f_c = \frac{1}{\exp\left(\frac{E_h - E_{f_p}}{kT}\right) + 1} - \frac{1}{\exp\left(\frac{E_e - E_{f_n}}{kT}\right) + 1} \tag{16}$$

And for intrinsic doping and equal effective masses, $E_{f_p} - E_i = -\frac{\Delta\mu}{2}$ and $E_{f_n} - E_i = \frac{\Delta\mu}{2}$ by symmetry arguments. Here, $E_{f_{p/n}}$ is the quasi-Fermi level for the holes/electrons, $E_i$ is the Fermi level of the intrinsic system (at mid-gap), and $\Delta\mu = E_{f_n} - E_{f_p}$ is the quasi-Fermi level splitting.

By symmetry of the electron and hole in this case, we must have $E_e - E_i = \frac{E}{2}$ and $E_h - E_i = -\frac{E}{2}$, where $E$ is the energy of the photon. Thus,

$$f_v - f_c = \frac{1}{\exp\left(-\frac{E - \Delta\mu}{2kT}\right) + 1} - \frac{1}{\exp\left(\frac{E - \Delta\mu}{2kT}\right) + 1} \tag{17}$$

For simplicity in analysis, let us set $x = \frac{E - \Delta\mu}{kT}$. Thus,

$$f_v - f_c = \frac{1}{e^{-\frac{x}{4}}\left(e^{-\frac{x}{4}} + e^{\frac{x}{4}}\right)} - \frac{1}{e^{\frac{x}{4}}\left(e^{\frac{x}{4}} + e^{-\frac{x}{4}}\right)} = \frac{\left[\exp\left(\frac{x}{4}\right) + \exp\left(-\frac{x}{4}\right)\right]\left[\exp\left(\frac{x}{4}\right) - \exp\left(-\frac{x}{4}\right)\right]}{\left[\exp\left(\frac{x}{4}\right) + \exp\left(-\frac{x}{4}\right)\right]^2}$$

$$= \frac{\left[\exp\left(\frac{x}{4}\right) - \exp\left(-\frac{x}{4}\right)\right]}{\left[\exp\left(\frac{x}{4}\right) + \exp\left(-\frac{x}{4}\right)\right]} = \frac{\sinh\left(\frac{x}{4}\right)}{\cosh\left(\frac{x}{4}\right)} = \tanh\left(\frac{x}{4}\right) = \tanh\left(\frac{E - \Delta\mu}{4kT}\right)$$

We have argued already above that $\tanh\left(\frac{E - \Delta\mu}{4kT}\right)$ should serve as a good approximation to $f_v - f_c$ for most systems and should capture the main physics of band filling. It may be modified to yield more accurate results in the case of high doping or a large mismatch between the electron and hole effective masses under the parabolic bands approximation. For the purposes of this work, let us proceed with the simple expression so that the luminescence becomes

$$S_{PL}(E, \Delta\mu) = \left(\frac{4\pi L}{h^3 c^2}\alpha(E, 0)E^2\right)\left(\frac{\tanh\left(\frac{E - \Delta\mu}{4kT}\right)}{\exp\left(\frac{E - \Delta\mu}{kT}\right) - 1}\right) \tag{18}$$

Where the term on the left is a sole function of $E$ and the term on the right includes both $E$ and $\Delta\mu$. Note that by taking $x = \frac{E - \Delta\mu}{kT}$, we have

$$\frac{\tanh\left(\frac{x}{4}\right)}{\exp(x) - 1} = \frac{\sinh\left(\frac{x}{4}\right)}{\cosh\left(\frac{x}{4}\right)(\exp(x) - 1)}$$

$$= \frac{1}{\exp(x) - 1}\left[\exp\left(\frac{x}{4}\right) - \exp\left(-\frac{x}{4}\right)\right] / \left[\exp\left(\frac{x}{4}\right) + \exp\left(-\frac{x}{4}\right)\right]$$

$$= \frac{1}{\exp(x) - 1}\left[\exp\left(\frac{x}{2}\right) - 1\right] / \left[\exp\left(\frac{x}{2}\right) + 1\right]$$

Finally, we use that $\exp(x) - 1 = \left[\exp\left(\frac{x}{2}\right) - 1\right]\left[\exp\left(\frac{x}{2}\right) + 1\right]$, so that

$$\frac{\tanh\left(\frac{x}{4}\right)}{\exp(x) - 1} = \frac{1}{\left(\exp\left(\frac{x}{2}\right) + 1\right)^2}$$

Let us double check that there are no singularities as $x \to 0$, since $\frac{\tanh(0)}{\exp(0)-1} = \frac{0}{0}$. To do so, we shall use L'Hôpital's rule, i.e.

$$\lim_{x\to c}\frac{f(x)}{g(x)} = \lim_{x\to c}\frac{f'(x)}{g'(x)}$$

With $f(x) = \tanh\left(\frac{x}{4}\right)$ and $g(x) = \exp(x) - 1$, giving $f'(x) = \frac{1}{4}\mathrm{sech}^2\left(\frac{x}{4}\right)$ and $g'(x) = \exp(x)$. Thus, $\lim_{x\to 0}\frac{\tanh\left(\frac{x}{4}\right)}{\exp(x)-1} = \frac{1}{4}$, so that there are no singularities and the luminescence can be rewritten as

$$S_{PL}(E,\Delta\mu) = \frac{\frac{4\pi L}{h^3c^2}\alpha(E,0)E^2}{\left(\exp\left(\frac{E-\Delta\mu}{2kT}\right)+1\right)^2} \tag{19}$$

Which is positive definite and is a good approximation for the luminescence with significant band tailing while explicitly including the band filling effects. Note that when $\frac{E-\Delta\mu}{kT} \gg 1$, we have

$$S_{PL}(E,\Delta\mu) \approx \frac{4\pi L}{h^3c^2}\alpha(E,0)E^2\exp\left(-\frac{E}{kT}\right)\exp\left(\frac{\Delta\mu}{kT}\right) \tag{20}$$

which recovers the expression without band filling contribution, suitable for low injection and sharp band edges and has been the standard expression used in most detailed balance analyses of solar cells. It is clear from Equation 19 that the luminescence spectra and radiative current will scale non-linearly with $\Delta\mu$. Furthermore, for $\alpha(E,0) \sim \exp\left(\frac{E-E_g}{\gamma}\right)$, as in the case of Urbach tails, we can take a derivative of the luminescence flux and find that the peak position will occur at

$$E_{PL}^{max} = \Delta\mu - 2kT\ln\left(\frac{1}{\frac{2kT}{E_{PL}^{max}}+\frac{kT}{\gamma}} - 1\right) \tag{21}$$

For $\gamma > kT$. A simpler but approximate solution can be found by taking $E_{PL}^{max} \gg kT$, and neglecting that term, so that

$$E_{PL}^{max} \approx \Delta\mu + 2kT\ln\left(\frac{kT}{\gamma - kT}\right) \tag{22}$$

Which shows that the luminescence peak depends directly on $\Delta\mu$, for $\gamma > kT$

**Section S4. Effects of band tails on $J - V$ characteristics**

While Equation 19 suggests a rather complex dependence of the band filling characteristics on current, we find that the $J - V$ characteristics can be well fitted to a modified diode expression in most cases:

$$J_{rad}(V) \sim J_0(\gamma, E_g) \exp\left(\frac{qV}{n_{eff}(\gamma, E_g)kT}\right) \tag{23}$$

In other words, the effect of band filling and band tails is to modify the recombination current prefactor $J_0$ and effective ideality factor $n_{eff}$, which manifest in the voltage loss as described in the main text and in section S6. Of particular interest is $n_{eff}$, which should be measurable in electroluminescence measurements, because non-radiative dark current occurs in parallel to the radiative dark current. Thus, we would expect the $n_{eff}$ estimated here in Figure S5 to be accurate even in systems far away from the radiative limit, as long as we measure the radiative current flux through voltage-dependent electroluminescence. We note that the calculated $n_{eff}$ for a-Si (assuming $E_g \sim 1.7\ eV$ and $\gamma \sim 50\ meV$) is around 1.7, which is quite similar to the value measured by Rau et al[10]. To get an approximate analytic expression for $n_{eff}$, we use Eqn. 19 and assume that $E^2$ varies slowly compared to the exponentials in the integrand and that we are in the weakly absorbing limit. Thus, $J_{rad}(V) \sim \int dE \exp\left(\frac{E-E_g}{\gamma}\right)\left(\exp\left(\frac{E-V}{2kT}\right)+1\right)^{-2}$ and with some rewriting, we find that $J_{rad}(V) \sim \int dx \exp\left(\frac{kTx+V-E_g}{\gamma}\right)\left(\exp\left(\frac{x}{2}\right)+1\right)^{-2} \sim \exp\left(\frac{V}{\gamma}\right)$. That is, we expect

$$n_{eff} \approx \frac{\gamma}{kT} \tag{24}$$

Which seems to hold somewhat well for small $\gamma$ just above $kT$, as observed in Figure S5. Furthermore, using the diode approximation from above, we can also calculate the modified fill factor expression as

$$FF(n_{eff}, V_{oc}) \approx \frac{\frac{qV_{oc}}{n_{eff}kT} - \ln\left(1 + \frac{qV_{oc}}{n_{eff}kT}\right)}{1 + \frac{qV_{oc}}{n_{eff}kT}} \tag{25}$$

which reduces the fill factor slightly compared to the case without band tails and is an additional efficiency loss mechanism.

## Section S5. Two bandgap model for band tails

To develop a simple picture for the apparent bandgap shift, voltage loss, and effects of band tailing, we use a simplistic model of the absorbance parametrized by two step functions. We will refer to this as the "two bandgap model", whose absorbance can be seen in Figure S7(a) and is given by:

$$A(E) = a_1\theta(E - E_{g,1})\theta(E_{g,2} - E) + a_2\theta(E - E_{g,2}) \tag{26}$$

Where $a_2 = 1$ and $E_{g,2} = 1.34\ eV$. The above model represents a simplistic picture of a system with band tails as it deviates from the Shockley-Queisser limit. Here, $E_{g,2}$ defines the absorption bandgap, $E_{g,1}$ is the lower bandgap that forms as a result of band tailing, and $a_1$ is the effective sub-gap absorption. We then calculate the typical photovoltaic figures of merit in Figure S7(b) while varying $\Delta E_g = E_{g,2} - E_{g,1}$ and $a_1$. The result is qualitatively similar to what is seen with a band tail (e.g. see Figure S2 for comparison), where the efficiency loss is essentially all in the voltage. Moreover, there is a specific transition point where the voltage loss is linear with the bandgap separation, dependent on the value of $a_1$. To see this, recall that $V_{oc} = \frac{kT}{q}\ln\left(\frac{J_{sc}}{J_0}+1\right)$, where $J_0 = \int A(E)S_{BB}(E)dE$ and $A(E)$ is given in Eqn. 26. The loss due to a lower bandgap $E_{g,1}$ is then

$$\Delta V_{oc} = \frac{kT}{q}\ln\left(\frac{J_{sc}}{J_2}+1\right) - \frac{kT}{q}\ln\left(\frac{J_{sc}}{J_0}+1\right) \approx -\frac{kT}{q}\ln\left(\frac{J_1}{J_2}+1\right) \tag{27}$$

Where $J_{1,2} = \int A_{1,2}(E)S_{BB}(E)dE$, $A_{1,2}(E) = a_{1,2}\theta(E - E_{g,1,2})$, and we have assumed $J_{sc} \gg J_0, J_2$. Thus, from the perspective of the voltage loss in the detailed balance analysis, $E_{g,1}$ does not appear as a photovoltaic bandgap until $J_1 > J_2$. This occurs when

$$\frac{a_1 e^{-\frac{E_{g,1}}{kT}}\left(\left(\frac{E_{g,1}}{kT}\right)^2 + 2\left(\frac{E_{g,1}}{kT}\right) + 2\right)}{a_2 e^{-\frac{E_{g,2}}{kT}}\left(\left(\frac{E_{g,2}}{kT}\right)^2 + 2\left(\frac{E_{g,2}}{kT}\right) + 2\right)} > 1 \tag{28}$$

Assuming $E_{g,1,2} \gg kT$, we can neglect the terms outside of the exponential to first order because it shows up logarithmically with $\Delta E_g$. Thus, the transition to a new bandgap occurs when

$$\Delta E_g > kT\ln\left(\frac{a_2}{a_1}\right) \tag{29}$$

In other words, from the perspective of the Shockley-Queisser limit and voltage loss, the Stokes shift is not apparent until Eqn. 29 is satisfied. At this point, the voltage loss scales linearly with increasing $E_{g,1}$. To see this clearly, we plot the voltage loss with bandgap shift with energies and voltages normalized to $kT\ln\left(\frac{a_2}{a_1}\right)$ in Figure S7(c). We see that indeed the transition occurs under the condition of Eqn. 29, where thereafter $\frac{\partial \Delta V_{oc}}{\partial \Delta E_g} \approx 1$. This is true irrespective of the value of $a_1$. Moreover, while Eqn. 29 is derived for two discrete bandgaps, we can generalize the concept to how sharp a continuous absorption spectrum should be to avoid a Stokes shifted voltage loss. Let us define $a_2 = \Delta a + a_1$ and take the limit as $\Delta a, \Delta E \to 0$. Thus, the generalized continuous form of Eqn. 29 becomes

$$\frac{kT}{a}\frac{\partial a}{\partial E} < 1 \tag{30}$$

In the case of weakly absorbing Urbach band tails, $a \sim \alpha L \sim C\exp\left(\frac{E-E_g}{\gamma}\right)$. Thus, Eqn. 30 predicts a Stokes shift should occur when $\gamma > kT$, which is what we observe in Figure 2 of the manuscript.

## Section S6. General Expression for Voltage Loss due to Nonabrupt Band Edges

The plots of Figure 3 and Figure S7 in the main text suggests a general relation between bandgap shifts and voltage loss, irrespective of the exact functional form of the band edge. To see this, note that the majority of the luminescence of the step-function absorbance is concentrated within $kT$ of the band edge and its integral varies exponentially with the bandgap energy. Thus, the effective bandgap of the luminescence, $E_{g,PL}$ must be chosen to integrate to nearly the majority of the luminescence flux. Thus, we pragmatically define it as

$$\max(E_{g,PL}) \ni \frac{\int_{E_{g,PL}}^{\infty} S_{PL}(E,\Delta\mu)dE}{\int_{0}^{\infty} S_{PL}(E,\Delta\mu)dE} \geq 0.90 \tag{31}$$

While this definition of $E_{g,PL}$ is not unique, it parametrizes the luminescence typically assumed under step-function absorbance to a greater variety of luminescence spectra and is somewhat less sensitive to noise. We further define the above-gap absorbance as

$$\bar{a}_{AG} = \frac{\int_{E_{g,Abs}}^{\infty} a(E,\Delta\mu)bb(E,\Delta\mu)dE}{\exp(\Delta\mu/kT)\int_{E_{g,Abs}}^{\infty} bb(E,0)dE} = \frac{\int_{E_{g,Abs}}^{\infty} S_{PL}(E,\Delta\mu)dE}{\exp(\Delta\mu/kT)\int_{E_{g,Abs}}^{\infty} bb(E,0)dE} \tag{32}$$

And below-gap absorbance as

$$\bar{a}_{SG} = \frac{\int_{E_{g,PL}}^{E_{g,Abs}} a(E,\Delta\mu)bb(E,\Delta\mu)dE}{\exp(\Delta\mu/kT)\int_{E_{g,PL}}^{E_{g,Abs}} bb(E,0)dE} = \frac{\int_{E_{g,PL}}^{E_{g,Abs}} S_{PL}(E,\Delta\mu)dE}{\exp(\Delta\mu/kT)\int_{E_{g,PL}}^{E_{g,Abs}} bb(E,0)dE} \tag{33}$$

Where both values are apparently dependent on $\Delta\mu$. Accurate estimation of the quantity $S_{PL}(E,\Delta\mu)/\exp(\Delta\mu/kT)$ can be achieved by taking $E \gg \Delta\mu$ and fitting the luminescence spectra to the high energy absorption/EQE, or by fitting the full spectrum with the band filling factor. Alternatively, since Eqn. 6 of the main text only requires knowledge of the ratio $\bar{a}_{AG}/\bar{a}_{SG}$, we can simply use the directly measured luminescence spectrum:

$$\frac{\bar{a}_{AG}}{\bar{a}_{SG}} = \frac{\int_{E_{g,Abs}}^{\infty} S_{PL}(E,\Delta\mu)dE}{\int_{E_{g,PL}}^{E_{g,Abs}} S_{PL}(E,\Delta\mu)dE} \frac{\int_{E_{g,PL}}^{E_{g,Abs}} bb(E,0)dE}{\int_{E_{g,Abs}}^{\infty} bb(E,0)dE} \tag{34}$$

And the definitions of $E_{g,Abs}$ and $E_{g,PL}$ to estimate the weighted absorbance ratio.

These definitions work well because the integrated number of recombination electrons is what matters in the detailed balance analysis, which is achieved by the appropriate definitions of weighted absorption and bandgaps. Therefore, the voltage loss is given by a form that is quite similar to Eqn. 27:

$$\Delta V_{oc} = \frac{kT}{q}\ln\left(\frac{\bar{a}_{SG}}{\bar{a}_{AG}}\exp\left(\frac{E_{g,Abs}-E_{g,PL}}{kT}\right)\left[\frac{\left(\frac{E_{g,PL}}{kT}\right)^2 + \frac{2E_{g,PL}}{kT} + 2}{\left(\frac{E_{g,Abs}}{kT}\right)^2 + \frac{2E_{g,Abs}}{kT} + 2}\right] + 1 - \frac{\bar{a}_{SG}}{\bar{a}_{AG}}\right) \tag{34}$$

Noting the logarithmic dependence on the argument and assuming $\Delta E_g = E_{g,Abs} - E_{g,PL} \ll E_{g,Abs}$, as well as $E_{g.Abs}, E_{g,PL} \gg kT$, we arrive at a simple expression that only depends on the observed bandgap shifts and the ratio of the above-gap and sub-gap absorbances:

$$\Delta V_{oc}\left(\frac{\bar{a}_{SG}}{\bar{a}_{AG}}, \Delta E_g\right) \approx \frac{kT}{q} \ln\left(\frac{\bar{a}_{SG}}{\bar{a}_{AG}} \exp\left(\frac{\Delta E_g}{kT}\right) + 1 - \frac{\bar{a}_{SG}}{\bar{a}_{AG}}\right) \tag{35}$$

Note that this expression recovers the expected values of voltage loss as $\frac{\bar{a}_{SG}}{\bar{a}_{AG}} \rightarrow 0,1$ and as $\Delta E_g \rightarrow 0$. Furthermore, the functional form of the sub-gap absorbance is captured by its effect of varying the value of $\bar{a}_{SG}$. From an experimental standpoint, another method to estimate the voltage loss is by using the modified $J - V$ characteristics found in Section S4. It is clear then that $J_{rad}(V) \approx J_{0,rad} \exp\left(\frac{qV}{n_{eff}kT}\right) = \int S_{EL}(E,V)dE$. Furthermore, it is possible to estimate $n_{eff}$ directly from the slope of voltage-dependent electroluminescence $S_{EL}(E,V)$. Integrating over $S_{EL}(E,V)$ and dividing by $\exp\left(\frac{qV}{n_{eff}kT}\right)$ then yields $J_{0,rad}$. Note that the $V_{oc}$ loss due to an imperfect band edge can be equivalently written in the form of $\Delta V_{oc} = \frac{\mathrm{kT}}{\mathrm{q}} \ln\left(\frac{J_{sc,SQ}}{J_{0,rad,SQ}}\right) - \frac{n_{eff}kT}{q} \ln\left(\frac{J_{sc}}{J_{0,rad}}\right)$ using Eqn. 23, which can be expanded to yield

$$\Delta V_{oc} = \frac{kT}{q} \ln\left(\frac{J_{sc,SQ}}{J_{sc}}\right) + \frac{kT}{q} \ln\left(\frac{J_{0,rad}}{J_{0,rad,SQ}}\right) - \frac{(n_{eff} - 1)kT}{q} \ln\left(\frac{J_{sc}}{J_{0,rad}}\right) \tag{36}$$

where the first term is the voltage loss due to incomplete absorption above the bandgap. The second term is the voltage loss due to band tailing, while the third term is a voltage *gain* due to band filling effects (e.g. see Fig. S4).

**Section S7. Effects of Sub-Unity Radiative and Quantum Efficiencies**

We have thus far only analyzed the situation assuming the reciprocity between absorption and photoluminescence, which holds quite generally but concerns primarily the internal open circuit voltage of a device i.e. the quasi Fermi level splitting. To analyze the effects of a system with sub-unity quantum efficiencies, which may be particularly relevant for localized states below the absorption gap, we assume Donolato's theorem still holds and apply Eqn. 9. Therefore, by reciprocity, the injection efficiency into these localized states would be relatively low, lowering the electroluminescence recombination rate and increasing the limiting $V_{oc}$ (see Figure S8). This situation would be analogous to considering free carrier absorption in the absorption band tail, where $IQE \rightarrow 0$, and therefore the absorption of free-carriers do not lead to photovoltaic current[11]. Thus, photogenerated carriers that do not contribute to photovoltaic current, whether they are localized states or free carriers, would not result in a loss to the open circuit voltage in the radiative limit. In general, the effect of band tails on the radiative limit should be determined via photocurrent spectroscopies, which captures this effect experimentally directly.

To analyze the voltage loss effects away from the radiative limit, i.e., sub-unity radiative efficiency, we note that generally Eqn. 23 holds and the discussion in section S4 suggests that $J(V) = J_{sc} - \frac{J_{0,rad}}{\eta_{ext}}\left(\exp\left(\frac{qV}{n_{eff}kT}\right) - 1\right)$ which is quite similar to Eqn. 11. Thus, it is readily apparent that the loss due to non-radiative recombination is modified with an ideality factor $n_{eff} \geq 1$, so that

$$\Delta V_{oc,nr} = -\frac{n_{eff}kT}{q}|\ln(\eta_{ext})|. \tag{37}$$

It should be noted that $\eta_{ext}$ is generally a function of voltage as well and should be measured/calculated at the operating voltage. This radiative ideality factor has already been recognized by Rau et al. to be relevant in amorphous Si[10] when analyzing its non-radiative losses. In many devices, sub-unity radiative efficiencies and sub-unity quantum efficiencies are both present and are likely competing to provide the observed voltage. In contrast, concentration benefits the voltage by a similar factor $\Delta V_{oc,conc} = \frac{n_{eff}kT}{q}|\ln(C)|$, where $C > 1$ is the concentration factor.

**Section S8. Parametrization of the Band edge Functional Form**

In Figure 3 of the main text, we considered various band edge functional forms to argue that there exists a general expression that relates $\Delta V_{oc,rad}$ to the existence of a Stokes shift, i.e. $\Delta E_g$. We considered two main types of band edges: exponential tails and indirect edge power laws. Exponential tails are the main form of band edges we have discussed in this article and we have thus far used the analysis described in Section S2. For the calculations in Figure S9 and Figure 3 of the main text, we vary $\theta$, $E_g$, and $\gamma$ to generate various functional forms for the band tail given by Eqn. 13. Furthermore, we consider only the absorption spectra that yield a luminescence bandgap above $4kT$, since we assume that $E_g \gg kT$ in deriving Eqn. 6 of the main text. We further consider a general power law form for a semiconductor band edge that has a weak oscillator strength (e.g. an indirect transition) with a higher energy direct transition with larger oscillator strengths:

$$\alpha(E) = \alpha_{0,ind}\left(E - E_{g,ind}\right)^n \theta\left(E - E_{g,ind}\right)\theta\left(E_{g,dir} - E\right) + \alpha_{0,dir}\theta\left(E - E_{g,dir}\right) \tag{39}$$

Where $E_{g,ind}$ and $E_{g,dir}$ represent the indirect and direct band edge, respectively, while $\alpha_{0,ind}$ and $\alpha_{0,dir}$ represent the absorption coefficients of the indirect and direct gaps, respectively. $n$ parametrizes the different energetic scaling relations of the indirect edge, typically $n < 3$ experimentally.

For both forms of band edges, we calculate $V_{oc,rad}$ from the complete modified detailed balance analysis, including band filling effects and assuming $\eta_{ext} = 1$ (Eqn. 11), $E_{g,Abs}$ is then derived from the calculated absorption spectrum using Rau's definition, and therefore $V_{oc,SQ}(E_{g,Abs})$ is calculated using a step-function at $E_{g,Abs}$. $E_{g,PL}$, $\bar{a}_{SG}$, $\bar{a}_{AG}$ is then calculated from the definitions in Section S6 by examining the luminescence spectra, $S_{PL}$. The results of these different band edges map well onto a simple relation described by Eqn. 35, suggesting a two bandgap model is an adequate representation of most experimentally observed band edge forms.

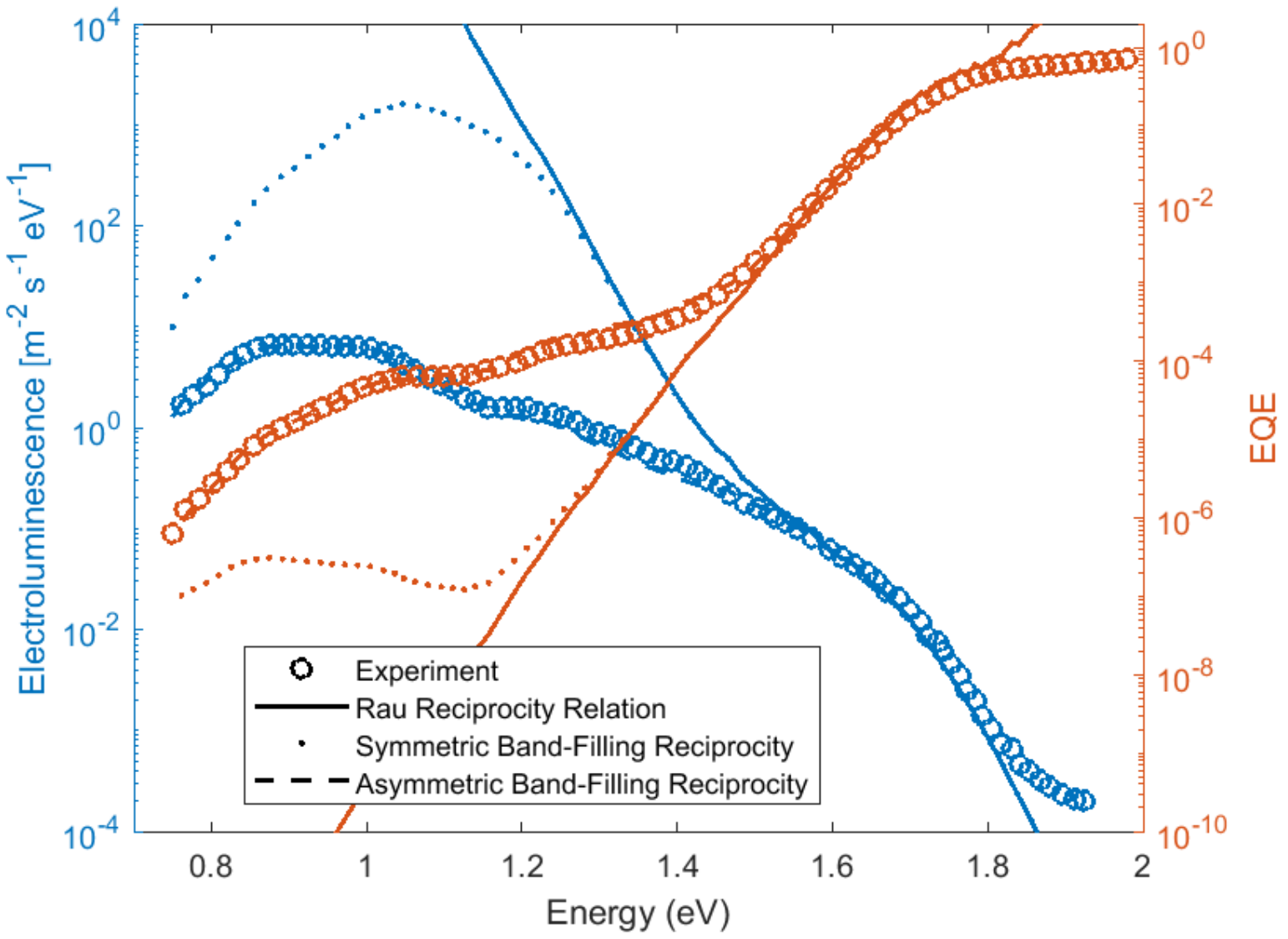

Figure S1. **Accounting for band filling in modified reciprocity relations**: Experimentally measured a-Si:H EQE and EL from ref. 10 (open circles). Solid lines correspond to the Rau reciprocity relation, whereas the dashed line is a fit that includes band filling effects with asymmetric effective masses in the parabolic approximation ($m_h/m_e = 1.818$, $\Delta\mu = 1.164\ V$, and $E_g = 2.439\ eV$. The dotted line includes band filling with the same fitted parameters except $m_h/m_e = 1$, i.e. assumes symmetric effective masses. All spectra are normalized by $\exp\,(\Delta\mu/kT)$ and the various reciprocity relations overlay for $E > E_{g,Abs}$.

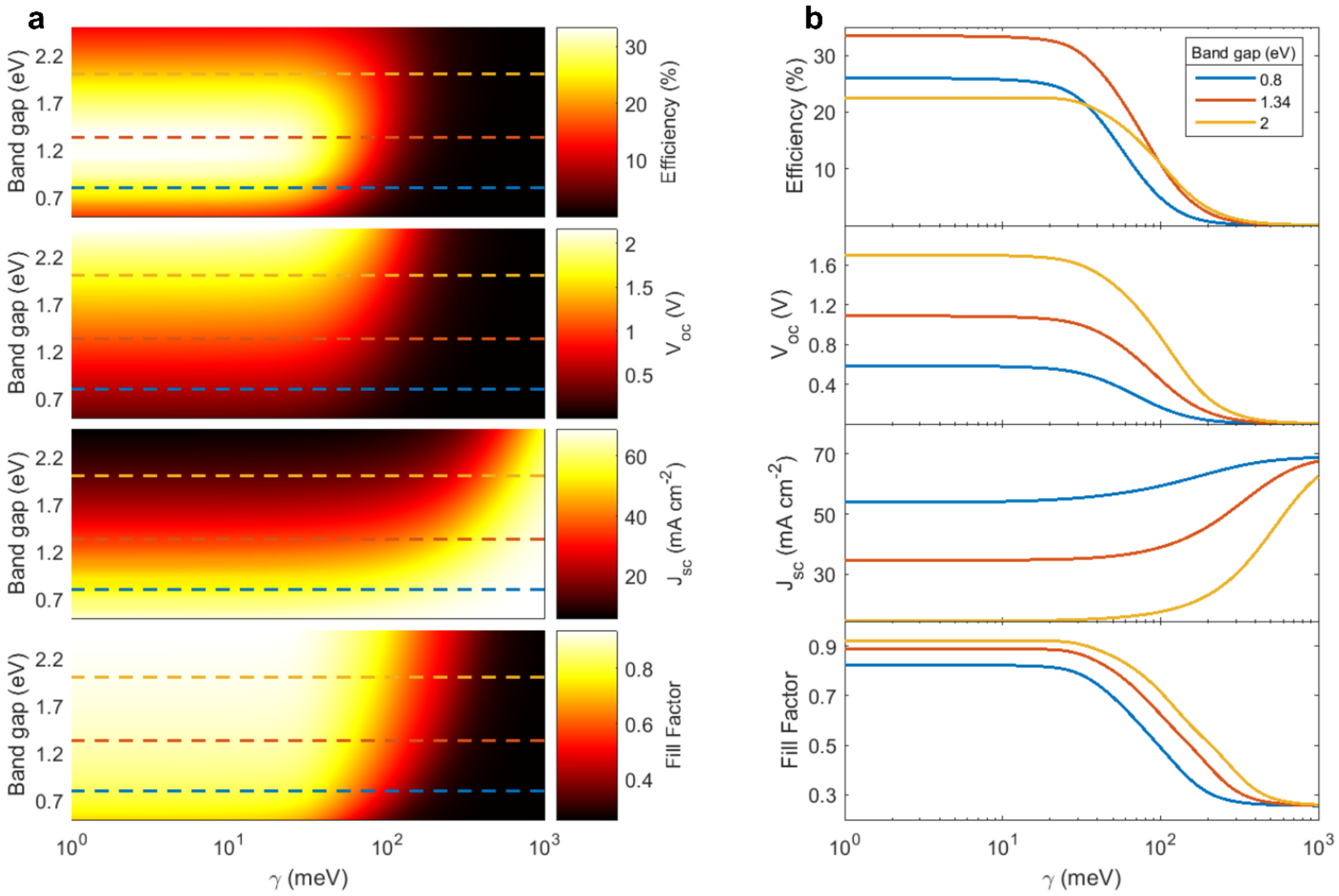

Figure S2. **Dependence of photovoltaic figures of merit on the Urbach parameter**: (a) The detailed balance limited value of conversion efficiency, open circuit voltage, short circuit current, and fill factor for different bandgaps and Urbach parameters assuming a thickness of $\alpha_0 L = 1$. (b) Linecuts of (a) at specific bandgap values.

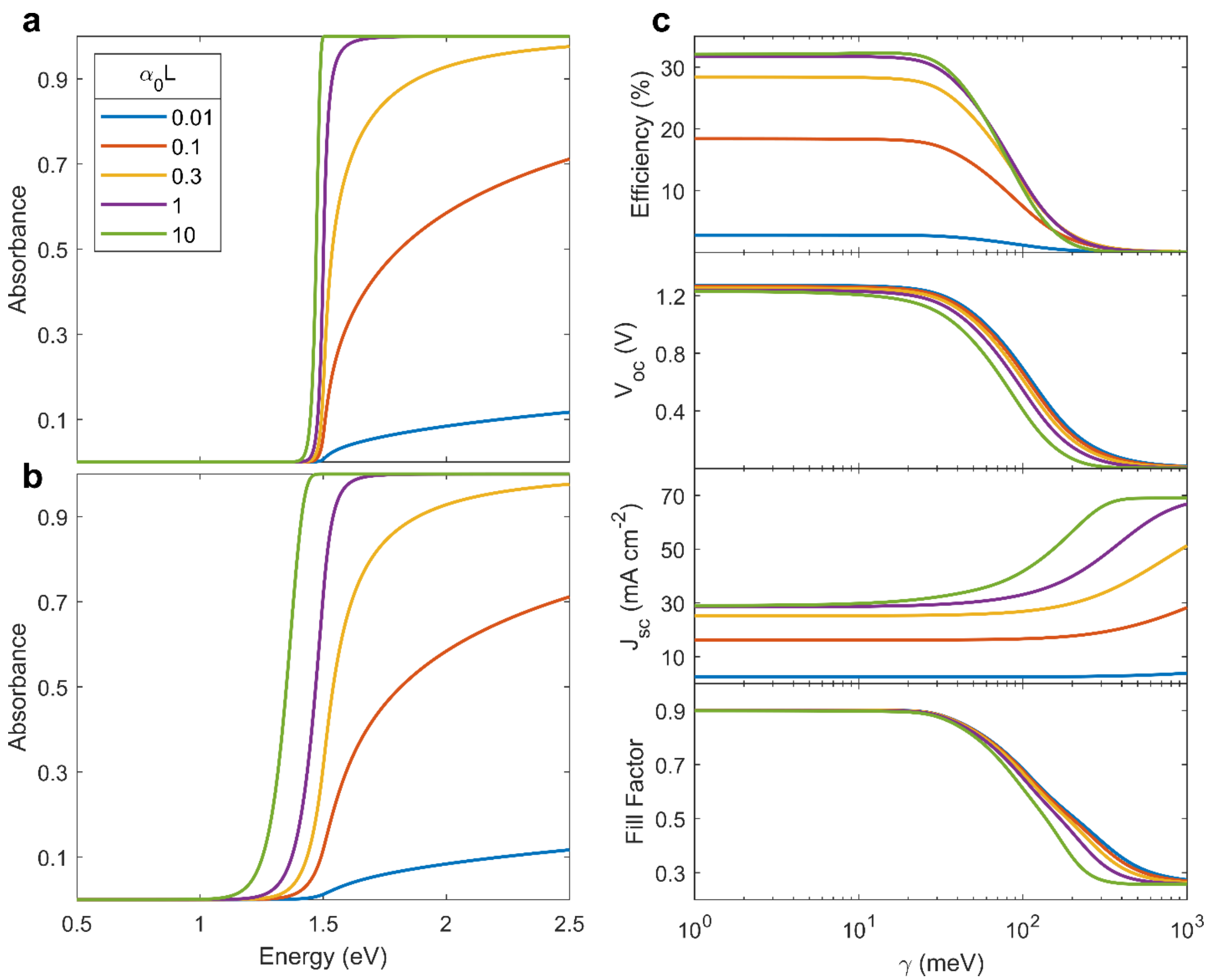

Figure S3. **Effects of thickness on photovoltaic figures of merit**: (a) Absorbance of a photovoltaic cell plotted with different normalized thicknesses ($\alpha_0 L$) for $\gamma = 0.5kT$ and (b) $\gamma = 2kT$ assuming a bandgap $E_g = 1.5$ eV. (c) Conversion efficiency, open circuit voltage, short circuit current, and fill factor calculated for different normalized thicknesses assuming a bandgap $E_g = 1.5$ eV. The different colored lines correspond to the same legend shown in (a).

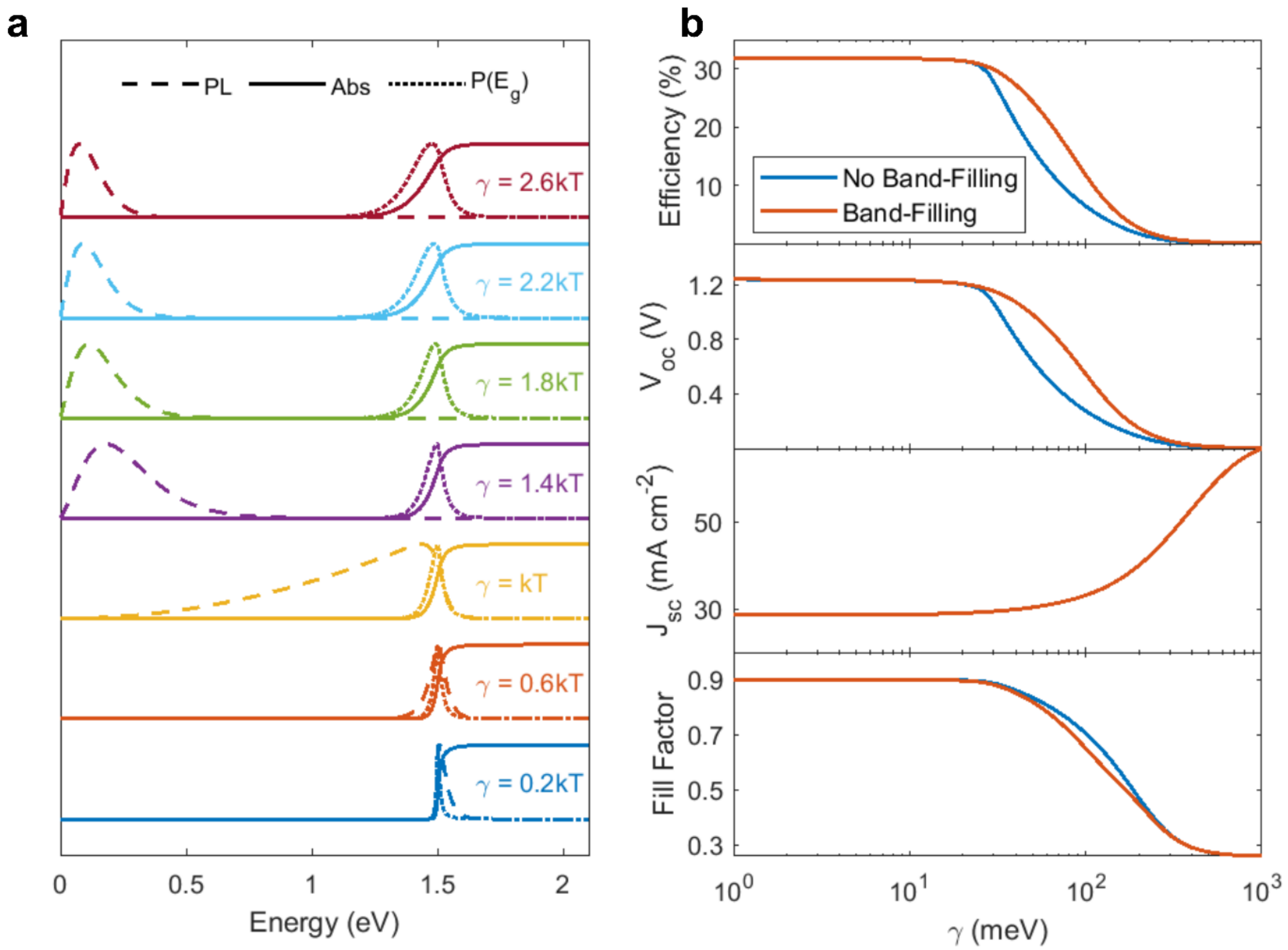

Figure S4. **The importance of including band filling effects**: (a) Calculated absorbance (solid line), photoluminescence (dashed line), and distribution of bandgaps (dotted line) for different Urbach parameters ($\gamma$) without including band filling effects. (b) Calculated efficiency, open circuit voltage, short circuit current, and fill factor with (orange solid line) and without (blue solid line) including band filling effects.

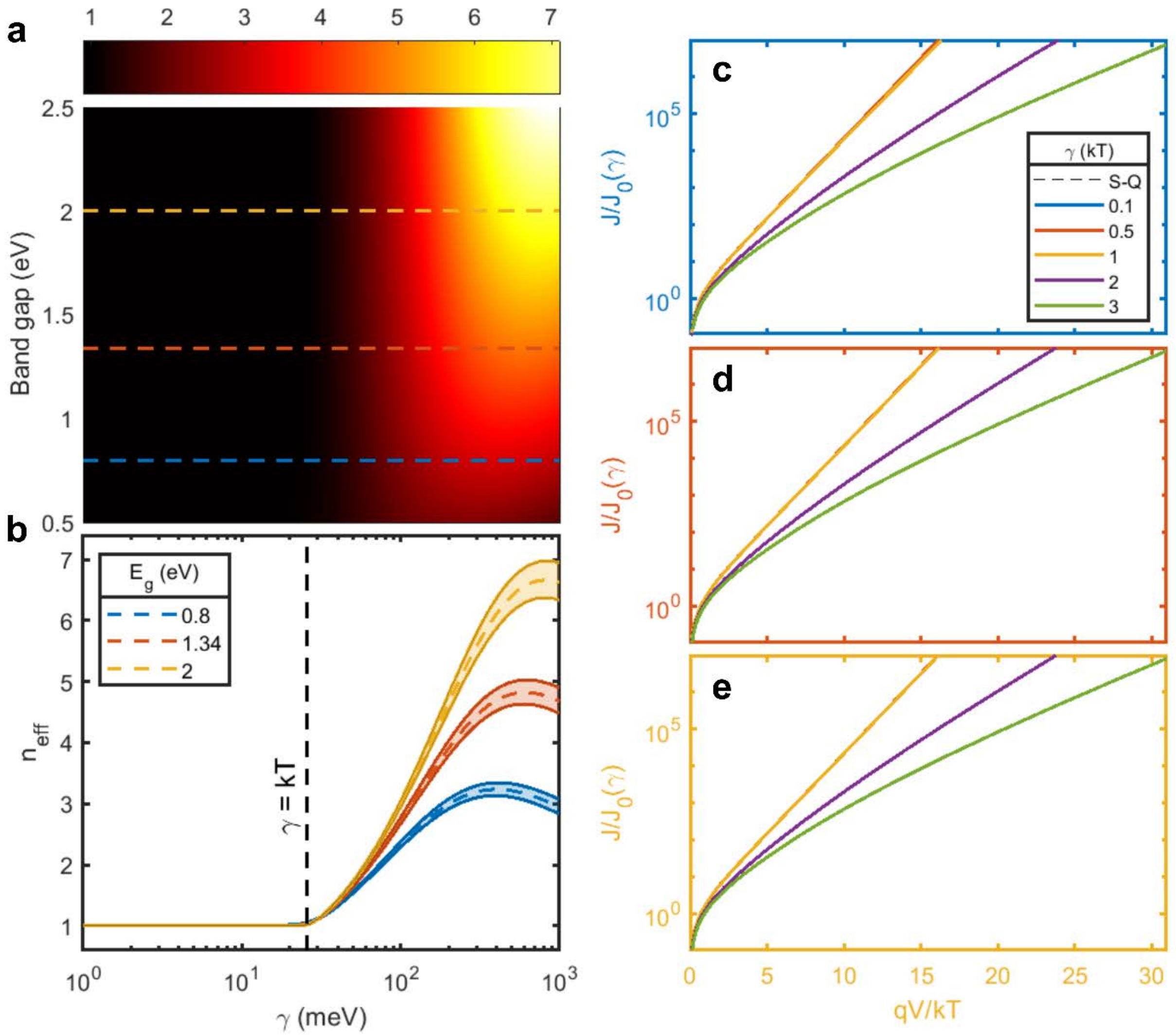

Figure S5. **Effects of band tails and band filling on ideality factor and current-voltage relationships**: (a) Fitted $n_{eff}$ for varying Urbach parameter ($\gamma$) and bandgap $E_g$. Fits were performed for the range $3kT < qV < E_g - 3kT$. Linecuts of (a) occur at $E_g = 0.8$ (blue), 1.34 (orange), and 2.0 eV (yellow). (b) Corresponding linecuts of (a) plotted for varying Urbach parameter ($\gamma$). Note the transition that occurs at $\gamma = kT$ to larger effective ideality factors, corresponding to the onset of band tailing and band filling effects. Dashed lines represent the fit, while solid lines represent the 95% confidence interval. $J - V$ characteristics for different bandgaps of 0.8 eV (c), 1.34 eV (d), and 2.0 eV (e). The different lines in a given plot represent different Urbach parameters. The legend in (c) is the same for (d) and (e). All plots have voltages normalized to $kT/q$ and current densities normalized to their radiative dark current $J_0$, which is a function of $\gamma$. Thicknesses were assumed to be $\alpha_0 L = 1$. Note that for Urbach parameters typically observed in experiment (i.e. $\gamma < \sim 3kT$), $n_{eff}$ is generally less than 3. For larger Urbach parameters, a modified ideality factor no longer describes the voltage scaling appropriately, since $E_{g,PL} \rightarrow kT$.

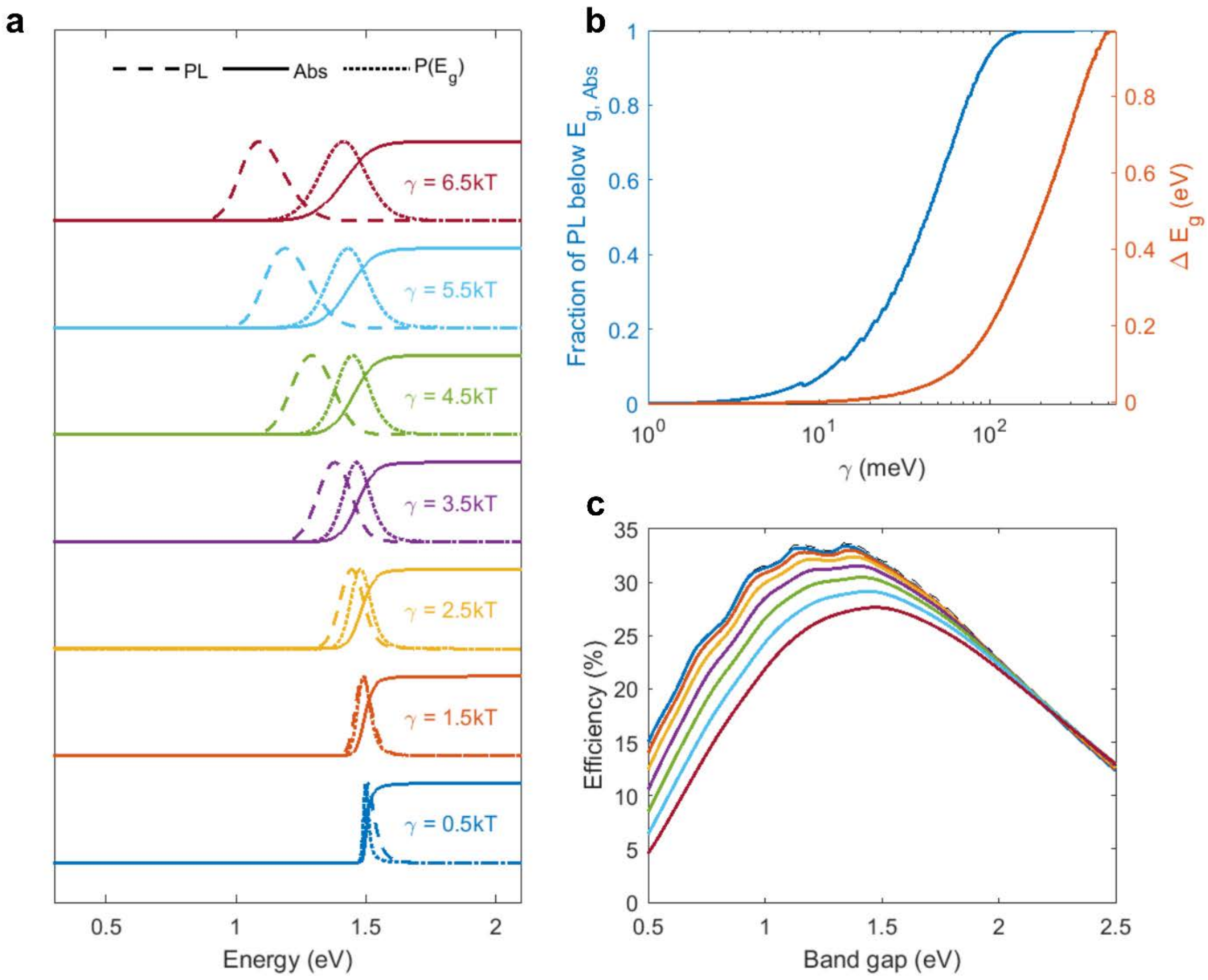

Figure S6. **Analysis of a Gaussian band tail distribution**: (a) Calculated absorbance (solid line), photoluminescence (dashed line) and distribution of bandgaps (dotted line) for an increasing Gaussian tail ($\gamma$). Here, the Gaussian tail distribution is calculated by taking $\theta = 2$ in Eqn. 14. (b) Fraction of integrated photoluminescence below the band gap (solid blue line) and Stokes shift $\Delta E_g$ (solid orange line) for a Gaussian tail distribution. (c) Calculated detailed balance efficiency for different bandgaps plotted for increasing Gaussian tail widths. The different colored lines correspond to the same values of the Gaussian tail displayed in (a).

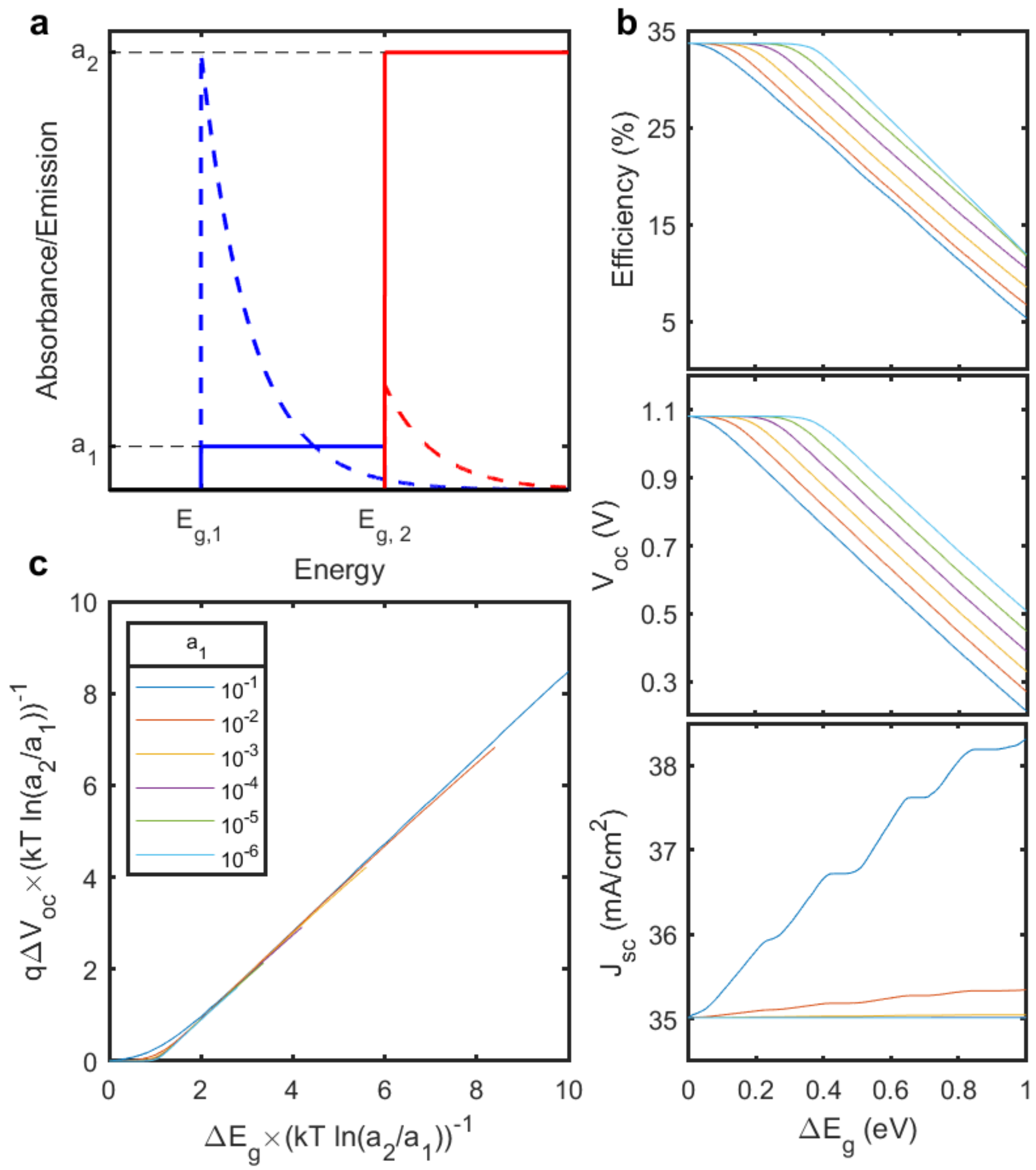

Figure S7. **Analysis of a two-bandgap toy model**: (a) Absorbance and emission of the two-bandgap toy model, parametrized by two step-functions. Solid lines correspond to absorbance, whereas dashed lines correspond to emission. (b) Plot of the photovoltaic figures of merit $(\eta, V_{oc}, J_{sc})$ for varying bandgap difference $\Delta E_g = E_{g,2} - E_{g,1}$ and values of the lower bandgap absorbance $a_1.a_2$ is assumed to be 1 while $E_{g,2} = 1.34$ eV. Colors correspond to the same as the legend in (c). (c) Voltage loss versus bandgap difference in normalized units of $kT \ln\left(\frac{a_2}{a_1}\right)$, showing the transition to the Stokes shift behavior for large enough band gap separation, dependent on $a_2/a_1$.

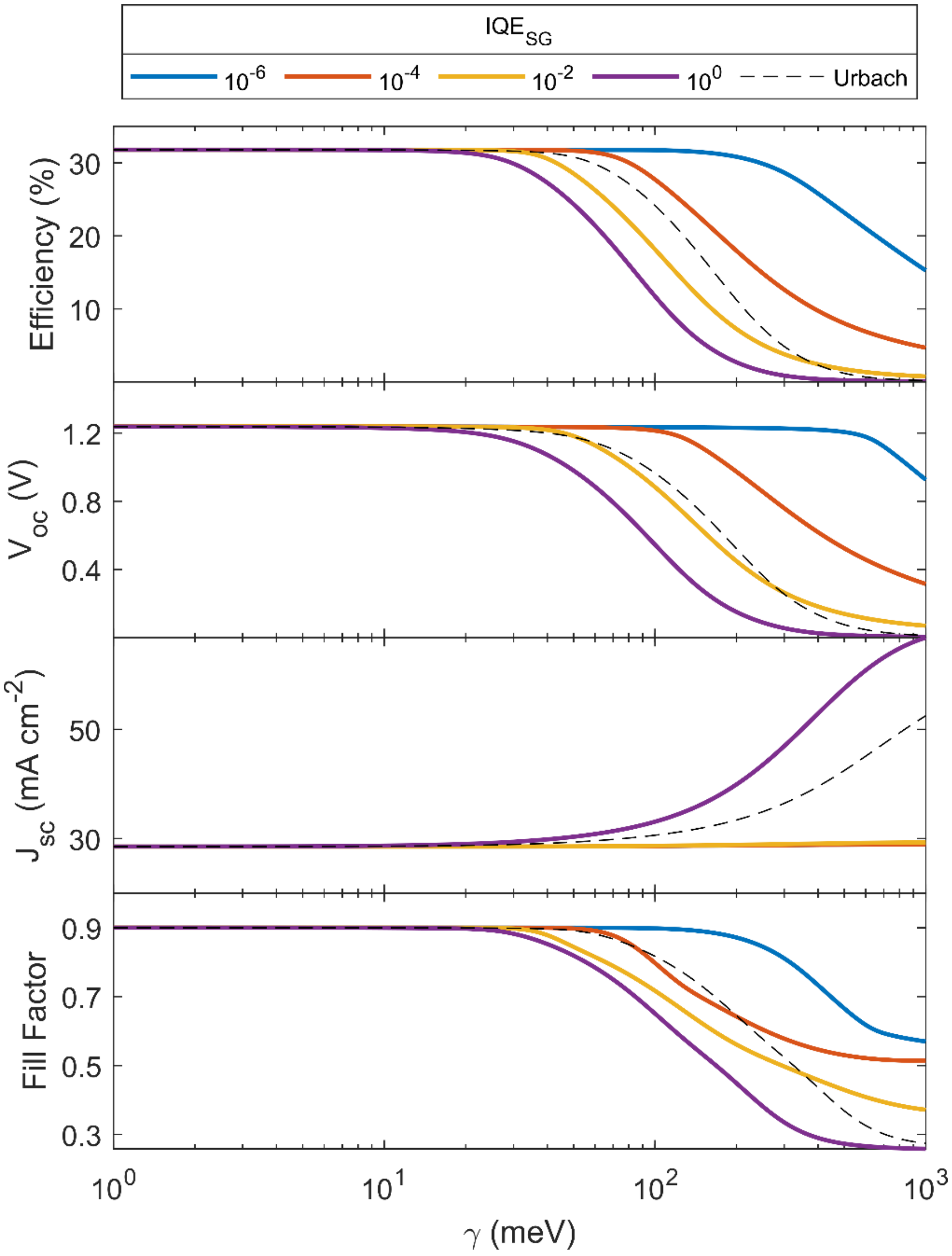

Figure S8. **Effects of a sub-unity collection efficiency below the bandgap**: Calculated power conversion efficiency, open circuit voltage, short circuit current density, and fill factor assuming that the collection efficiency below the bandgap ($IQE_{SG}$) is less than 1 and given by a constant average value. That is, we take the external quantum efficiency to be $EQE(E) = a(E)\left(IQE_{SG}\theta\left(E_g - E\right) + \theta\left(E - E_g\right)\right)$. The "Urbach" curve is calculated assuming the collection efficiency decays with a similar Urbach parameter to that used in the absorption calculation (i.e. $IQE_{SG}(\gamma, E) = \exp\left(\frac{E-E_g}{\gamma}\right)$), which may approximate the mobility-edge better than a constant.

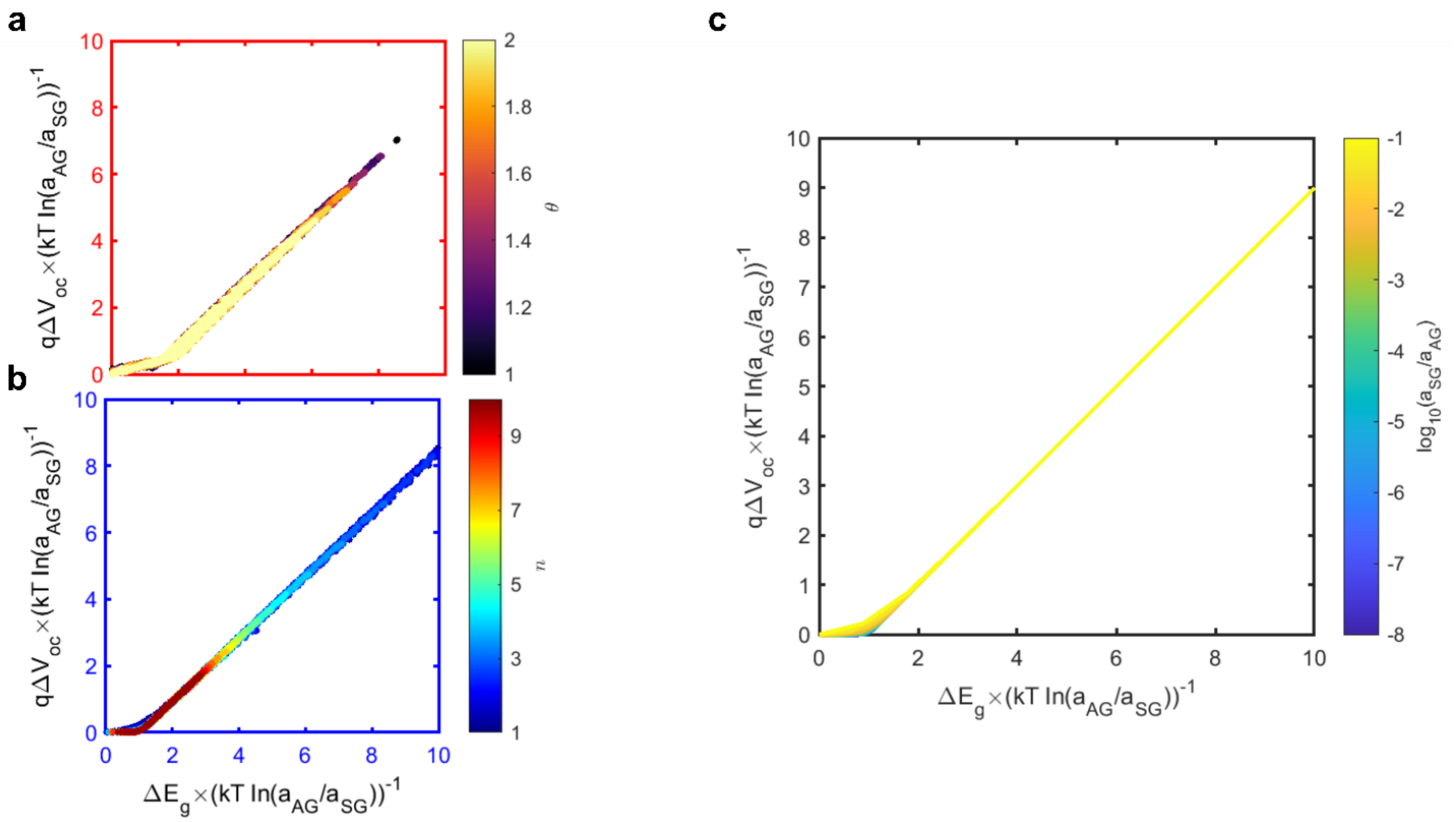

Figure S9. **Different band edges that map onto a two-bandgap model**: Stokes shift $\Delta E_g$ and radiative voltage loss $\Delta V_{oc}$ calculated from the full detailed balance analysis with the appropriate definitions of $E_{g,Abs}, E_{g,PL}, a_{AG}, a_{SG}$, as described in Section S6. We vary the parameters for the exponential band tail model (a) and the indirect edge power law model (b). For the exponential band tail model we take $\alpha_0 L = 10$, whereas for the indirect edge model we take $\alpha_{0,dir}L = 100$, $\alpha_{0,ind}L = 0.1$. Both forms map well onto the generalized expression (c). The colorbar for the generalized expression in (c) is $\log_{10}\left(\frac{a_{SG}}{a_{AG}}\right)$, i.e. describes the ratio of the sub-gap to above-gap absorption. The different ratios plots are overlaid, showing the similarity irrespective of $a_{SG}/a_{AG}$, assuming it is sufficiently small.

**Table S1**

| Material | Bandgap (eV) | Urbach Energy (meV) | Calculated $\Delta V_{oc}$ (mV) | References |
|---|---|---|---|---|
| c-Si | 1.12 | 9.6 | 24.0 | Cody 1992[12] |
| c-Si | 1.12 | 8.6 | 20.7 | Cody 1992[12] |
| c-Si | 1.12 | 11 | 28.9 | Green 2008[13] |
| GaAs | 1.42 | 6.9 | 14.6 | Sturge 1962[14] |
| GaAs | 1.42 | 7.5 | 16.3 | Johnson 1995[15] |
| GaAs | 1.42 | 5.9 | 11.7 | Beaudoin 1997[16] |
| InP | 1.355 | 9.4 | 22.5 | Subashiev 2010[17] |
| InP | 1.361 | 10.6 | 26.6 | Subashiev 2010[17] |
| InP | 1.34 | 7.1 | 16.3 | Beaudoin 1997[16] |
| a-Si:H | 1.72 | 42 | 273.1 | Cody 1984[18] |
| a-Si:H | 1.64 | 52 | 382.8 | Cody 1984[18] |
| a-Si:H | 1.69 | 67 | 559.1 | Cody 1981[19] |
| a-Si:H | 1.7 | 43 | 283.5 | Tiedje 1981[20] |
| a-Si:H | 1.69 | 47 | 329.0 | Tiedje 1983[21] |
| a-Si:H | 1.7 | 48 | 341.3 | van Veen 2003[22] |
| a-Si:H | 1.8 | 51 | 385.3 | van Veen 2003[22] |
| a-Si:H | 1.85 | 51 | 389.1 | van Veen 2003[22] |
| CdTe | 1.45 | 17 | 52.8 | Rakhshani 2001[23] |
| CdTe | 1.5 | 7.2 | 15.5 | Marple 1966[24] |
| CdTe | 1.5 | 9 | 21.1 | Mullins 1997[25] |
| CdTe | 1.5 | 10.6 | 26.5 | Sculfort 1984[26] |
| CIGS | 1.53 | 24 | 94.2 | Heath 2002[27] |
| CIGS | 1 | 18 | 56.0 | Heath 2002[27] |
| CIGS | 1.18 | 23 | 84.9 | Heath 2002[27] |
| CIGS | 1.2 | 31 | 143.6 | Troviano 2011[28] |
| CIGS | 1.67 | 25 | 102.5 | Meeder 2002[29] |
| CIGS | 1.08 | 9 | 21.9 | Shioda 1996[30] |
| Kesterite | 1.5 | 69 | 551.4 | Islam 2015[31] |
| Kesterite | 1.1 | 54 | 346.9 | Islam 2015[31] |
| Kesterite | 1.38 | 45 | 286.6 | Yan 2017[32] |
| Kesterite | 1.54 | 65 | 516.7 | Yan 2017[32] |
| Kesterite | 1.68 | 56.8 | 441.8 | Ng 2017[33] |
| Perovskite | 1.57 | 15 | 44.7 | De Wolf 2014[34] |
| Perovskite | 2.23 | 23 | 90.2 | Sadhanala 2014[35] |
| Perovskite | 1.57 | 14 | 40.5 | Zhang 2015[36] |
| Perovskite | 1.57 | 14.4 | 42.2 | Zhang 2015[36] |
| Perovskite | 1.57 | 15.8 | 48.3 | Zhang 2015[36] |
| Organic | 1.66 | 37 | 214.9 | Gotoh 1997[37] |
| Organic | 2 | 50 | 386.8 | Kronemeijer 2014[38] |
| Organic | 1.31 | 25.6 | 104.8 | Liu 2020[39] |
| Organic | 1.47 | 27 | 115.9 | Ran 2016[40] |
| Organic | 1.88 | 36 | 211.0 | Vandewal 2014[41] |
| Organic | 1.71 | 27 | 118.7 | Liu 2016[42] |
| Organic | 1.67 | 24 | 95.2 | Qian 2018[43] |